\documentclass[twocolumn, twocolappendix]{aastex631}

\shorttitle{Outflows in Medium-Band Images}
\shortauthors{Zhu et al.}
\newcommand{\cut}[1]{}
\newcommand{\revision}[1]{#1}
\newcommand{\rrevision}[1]{#1}

\begin{document}

\title{A Systematic Search for Galaxies with Extended Emission Line and Potential Outflows \\ in JADES Medium-Band Images}

\author[0000-0003-3307-7525]{Yongda Zhu}
\affiliation{Steward Observatory, University of Arizona, 933 North Cherry Avenue, Tucson, AZ 85721, USA}

\author[0000-0002-7893-6170]{Marcia J. Rieke}
\affiliation{Steward Observatory, University of Arizona, 933 North Cherry Avenue, Tucson, AZ 85721, USA}

\author[0000-0001-7673-2257]{Zhiyuan Ji}
\affiliation{Steward Observatory, University of Arizona, 933 North Cherry Avenue, Tucson, AZ 85721, USA}

\author[0000-0003-4770-7516]{Charlotte Simmonds}
\affiliation{Kavli Institute for Cosmology, University of Cambridge, Madingley Road, Cambridge, CB3 0HA, UK}
\affiliation{Cavendish Laboratory, University of Cambridge, 19 JJ Thomson Avenue, Cambridge, CB3 0HE, UK}

\author[0000-0002-4622-6617]{Fengwu Sun}
\affiliation{Center for Astrophysics $|$ Harvard \& Smithsonian, 60 Garden St., Cambridge MA 02138 USA}

\author[0000-0001-6561-9443]{Yang Sun}
\affiliation{Steward Observatory, University of Arizona, 933 North Cherry Avenue, Tucson, AZ 85721, USA}

\author[0000-0002-8909-8782]{Stacey Alberts}
\affiliation{Steward Observatory, University of Arizona, 933 North Cherry Avenue, Tucson, AZ 85721, USA}

\author[0000-0003-0883-2226]{Rachana Bhatawdekar}
\affiliation{
European Space Agency (ESA), European Space Astronomy Centre (ESAC), Camino Bajo del Castillo s/n, 28692 Villanueva de la Cañada, Madrid, Spain}

\author[0000-0002-8651-9879]{Andrew J.\ Bunker }
\affiliation{Department of Physics, University of Oxford, Denys Wilkinson Building, Keble Road, Oxford OX1 3RH, UK}

\author[0000-0002-1617-8917]{Phillip A.\ Cargile}
\affiliation{Center for Astrophysics $|$ Harvard \& Smithsonian, 60 Garden St., Cambridge MA 02138 USA}

\author[0000-0002-6719-380X]{Stefano Carniani}
\affiliation{Scuola Normale Superiore, Piazza dei Cavalieri 7, I-56126 Pisa, Italy}

\author[0000-0002-2380-9801]{Anna de Graaff}
\affiliation{Max-Planck-Institut f\"ur Astronomie, K\"onigstuhl 17, D-69117, Heidelberg, Germany}

\author[0000-0003-4565-8239]{Kevin Hainline}
\affiliation{Steward Observatory, University of Arizona, 933 North Cherry Avenue, Tucson, AZ 85721, USA}

\author[0000-0003-4337-6211]{Jakob M.\ Helton}
\affiliation{Steward Observatory, University of Arizona, 933 North Cherry Avenue, Tucson, AZ 85721, USA}

\author[0000-0002-0267-9024]{Gareth C.\ Jones}
\affiliation{Department of Physics, University of Oxford, Denys Wilkinson Building, Keble Road, Oxford OX1 3RH, UK}

\author[0000-0002-6221-1829]{Jianwei Lyu}
\affiliation{Steward Observatory, University of Arizona,
933 North Cherry Avenue, Tucson, AZ 85721, USA}

\author[0000-0003-2303-6519]{George H. Rieke}
\affiliation{Steward Observatory, University of Arizona, 933 North Cherry Avenue, Tucson, AZ 85721, USA}

\author[0000-0002-5104-8245]{Pierluigi Rinaldi}
\affiliation{Steward Observatory, University of Arizona, 933 North Cherry Avenue, Tucson, AZ 85721, USA}

\author[0000-0002-4271-0364]{Brant Robertson}
\affiliation{Department of Astronomy and Astrophysics University of California, Santa Cruz, 1156 High Street, Santa Cruz CA 96054, USA}

\author[0000-0001-6010-6809]{Jan Scholtz}
\affiliation{Kavli Institute for Cosmology, University of Cambridge, Madingley Road, Cambridge, CB3 0HA, UK}
\affiliation{Cavendish Laboratory, University of Cambridge, 19 JJ Thomson Avenue, Cambridge, CB3 0HE, UK}

\author[0000-0003-4891-0794]{Hannah \"Ubler}
\affiliation{Kavli Institute for Cosmology, University of Cambridge, Madingley Road, Cambridge, CB3 0HA, UK}
\affiliation{Cavendish Laboratory, University of Cambridge, 19 JJ Thomson Avenue, Cambridge, CB3 0HE, UK}

\author[0000-0003-2919-7495]{Christina C.\ Williams}
\affiliation{NSF’s National Optical-Infrared Astronomy Research Laboratory, 950 North Cherry Avenue, Tucson, AZ 85719, USA}

\author[0000-0001-9262-9997]{Christopher N.\ A.\ Willmer}
\affiliation{Steward Observatory, University of Arizona, 933 North Cherry Avenue, Tucson, AZ 85721, USA}

\correspondingauthor{Yongda Zhu}
\email{yongdaz@arizona.edu}

\begin{abstract}
 \revision{For the first time, we present a systematic search for galaxies with extended emission line and potential outflow features using \textit{JWST} medium-band images in the GOODS-S field. This is done by comparing the morphology in medium-band images to adjacent continuum and UV bands.} We look for galaxies that have a maximum extent 50\% larger, an excess area 30\% greater, or an axis ratio difference of more than 0.3 in the medium band compared to the reference bands. After visual inspection, we find 326 candidate galaxies at  \revision{$1.4 < z < 8.4$}, with a peak in the population near cosmic noon, benefiting from the good coverage of the medium-band filters. By \cut{examining}  \revision{fitting} their SEDs, we find that the candidate galaxies are at least 20\% more bursty in their star-forming activity and have \cut{60\%}  \revision{50\%} more young stellar populations compared to a control sample selected based on the continuum band flux. Additionally, these candidates exhibit a significantly higher production rate of ionizing photons. We further find that candidates hosting known AGN produce extended emission that is more anisotropic compared to non-AGN candidates. A few of our candidates have been spectroscopically confirmed to have prominent outflow signatures through NIRSpec observations, showcasing the robustness of the photometric selection. Future spectroscopic follow-up will better help verify and characterize the kinematics and chemical properties of these systems.
\end{abstract}

\keywords{
\href{http://astrothesaurus.org/uat/734}{High-redshift galaxies (734)}, 
\href{http://astrothesaurus.org/uat/572}{Galactic winds (572)},
\href{http://astrothesaurus.org/uat/582}{Galaxy classification systems (582)}
}

\section{Introduction} \label{sec:intro}
Understanding the origins of galaxy outflows and extended emission lines is crucial for unraveling the complex processes governing galaxy formation and evolution. Galactic outflows, driven by various mechanisms such as stellar winds, supernovae, and active galactic nuclei (AGN), play a significant role in regulating star formation by expelling gas and metals from galaxies \citep{veilleux_galactic_2005, somerville_physical_2015, naab_theoretical_2017, perrotta_kinematics_2023}. These outflows can also contribute to enriching the interstellar medium (ISM), the circumgalactic medium (CGM), and the intergalactic medium (IGM) \citep[e.g.,][]{weiner_ubiquitous_2009,tumlinson_circumgalactic_2017}. Without mechanisms like outflows to balance gas accretion and star formation, models predict much higher stellar-to-baryon ratios than observed \citep{white_core_1978, ubler_why_2014, somerville_physical_2015,naab_theoretical_2017,henriques_origin_2019}. Additionally, understanding the interplay between AGN-driven galactic outflows and AGN activities provides insights into the co-evolution of galaxies and their central supermassive black holes \citep{kormendy_coevolution_2013}. Recent studies have shown that cold gas outflows can significantly affect a galaxy's evolution \citep[e.g.,][]{veilleux_cool_2020,mingozzi_outflows_2021}, helping us piece together the life cycles of galaxies and their evolutionary pathways across cosmic time.

Significant progress has been made in understanding galactic outflows and extended emission lines—potential indicators of outflows—through both simulations and spectroscopic observations.  \revision{We caution that however, extended emission-line regions are not necessarily indicative of outflows. Such features can also arise from alternative physical processes, such as inflows, tidal stripping (as seen in jellyfish galaxies; \citealp{gondhalekar_systematic_2024}), or circumgalactic gas ionized by sources within the galaxy (e.g., extended Lyman-$\alpha$ halos; \citealp{peng_direct_2025}), as well as transient phenomena such as tidal disruption events \citep{wevers_extended_2024}.} State-of-the-art cosmological simulations, such as IllustrisTNG \citep{pillepich_first_2018,nelson_first_2019} and EAGLE \citep{crain_eagle_2015}, have provided valuable insights into the feedback processes that drive outflows and their impact on galaxy evolution. These feedback processes, driven by supernova explosions and AGN radiation pressure, are crucial for self-regulating star formation in galaxies \citep{debuhr_galaxy-scale_2012,ceverino_firstlight_2018,pandya_characterizing_2021}. Spectroscopic observations, particularly with integral-field spectrographs (IFS) on 8–10 m class telescopes as well as with the Atacama Large Millimeter/submillimeter Array (ALMA), have enabled detailed studies of the kinematics and chemical properties of outflows in both local and high-redshift galaxies \citep[e.g.,][]{spilker_fast_2018,forster_schreiber_kmos3d_2019,ginolfi_alpine-alma_2020,butler_molecular_2023}.

Besides spectroscopic observations \citep[e.g.,][]{weiner_ubiquitous_2009}, imaging data can also be used for studying outflows by tracing emission structures produced by outflow-ISM/CGM interaction. The Hubble Space Telescope (HST) has been crucial in identifying outflow candidates and studying their properties. High-resolution imaging observations from HST have revealed the presence of outflows in various forms, from large-scale ionized gas structures to more compact emission features \citep{heckman_absorption-line_2000}. The Cosmic Origins Spectrograph (COS) on HST has also provided valuable ultraviolet (UV) spectroscopic data, allowing for the study of outflow velocities and the physical conditions within these outflows \citep{bordoloi_cos-dwarfs_2014}. \citet{chisholm_shining_2016} highlighted the role of ionized outflows in shaping galaxy evolution, while their subsequent study in \citet{chisholm_galaxies_2017} discussed the connection between extreme outflows and ionizing photon leakage. Additionally, \citet{keel_hst_2015} used narrow- and medium-band HST imaging, along with ground-based imaging and spectra, to study fading AGN and their host galaxies, revealing important information about extended gas structures. These observations have established a foundation for understanding the prevalence and characteristics of galactic outflows, particularly in the local Universe. So far, however, outflow candidates based on extended emission  \revision{line} features have been discovered serendipitously, and there has not been a systematic search through deep  \revision{NIRCam} medium-band images.

The James Webb Space Telescope (JWST; \citealp{gardner_james_2023}), through its Near Infrared Camera (NIRCam; \citealp{rieke_performance_2023}) medium-band images, offers unprecedented opportunities to identify and study galaxies with potential outflow features across cosmic time. The medium-band filters provide a unique capability to capture emission lines from ionized gas, which may be indicative of outflows and extended emission features. Previous studies, such as those by \citet{keel_hst_2015} using HST, have shown the power of medium-band imaging in identifying extended gas structures. In this work, we utilize medium-band images from the JWST Advanced Deep Extragalactic Survey (JADES; \citealp{eisenstein_overview_2023,bunker_jades_2024}) and the JWST
Extragalactic Medium-band Survey (JEMS; \citealp{williams_jems_2023}) to identify galaxies in the Great Origins Deep Survey South (GOODS-S; \citealp{giavalisco_great_2004}) field with potential outflows by tracing extended emission line features. By comparing the morphology in medium-band images to adjacent continuum and UV bands, we aim to uncover a diverse population of galaxies exhibiting these phenomena. 

Our study presents the first systematic search for outflow candidates across cosmic time based on deep NIRCam medium-band imaging, serving as a pathfinder. This work complements and extends work by e.g., \citet{carniani_jades_2024}, that uses NIRSpec Multi-Object Spectroscopy (MOS) with Micro Shutter Assembly (MSA; \citealp{ferruit_near-infrared_2022}) spectra to study 52 low-mass star-forming galaxies ($M_*< 10^{10} M_\odot$) at $z > 3$ \citep[also see e.g., ][for other recent work based on NIRSpec/MOS data]{tang_jwstnirspec_2023,xu_stellar_2023,zhang_statistics_2024}. While \citet{carniani_jades_2024} identify potential ionized outflows traced by H$\alpha$ and/or [O III] emission lines through high-spectral-resolution observations, our approach leverages all the available medium-band imaging in JADES and JEMS GOODS-S to identify a broader sample of outflow candidates. In addition, NIRSpec/MOS studies have only focused on the outflow velocity and no (or little) information about the spatial extension has been reported. Therefore, NIRCam images can provide complementary information on the outflow properties. 
% Benefiting from the superior sensitivity and resolution of NIRCam, along with the multi-band photometry data, we aim to enhance our understanding of galaxy evolution through cosmic history.

The paper is structured as follows. We describe the data and methods we use in Section \ref{sec:data}. An overview of the selection results is presented in Section \ref{sec:results}. We then discuss the properties of the sample members in detail, including their comparison with the control sample, correlation between the extent of extended emission, the contribution of AGN, and spectroscopic observations, in Section \ref{sec:discussion}. Finally, we summarize the findings in Section \ref{sec:summary}. Throughout this paper, we use a flat $\Lambda$CDM cosmology with $\Omega_m = 0.315$ and $H_0=67.4$ km\ s$^{-1}$ \citep{planck_collaboration_planck_2020}. Distances and scales are quoted in proper units unless otherwise noted.

\section{Data and Selection Methods}\label{sec:data}

The idea behind our selection process is to identify galaxy candidates that exhibit larger extents in their medium-band images, which cover the emission lines, compared to the adjacent reference band --- typically the broad band to the red ---that primarily captures the stellar continuum. 
% We also ensure that the selected galaxies have larger extents than in the rest-UV band, to mitigate the risk of false detections. 
 \revision{Additionally, we ensure that the selected galaxies have larger extents than in the bluer band, which probes the rest-UV, to minimize the risk of misidentification.}
We focus our search in the GOODS-S field given the good coverage of NIRCam medium-band data and legacy ancillary data.

\subsection{Imaging data}
We use the JADES GOODS-S images from the initial data release \citep{rieke_jades_2023}, the JADES Origins Field (JOF; \citealp{eisenstein_jades_2023}), and the JWST Extragalactic Medium-band Survey (JEMS; \citealp{williams_jems_2023}). We use photometric redshifts determined by {\tt EAzY} \citep{brammer_eazy_2008} in \citet{hainline_cosmos_2024}. 
Medium-band filters used in this work include F182M, F210M, F250M, F300M, F335M, and F410M, which provide a redshift coverage for \cut{H$\alpha$ and [O III]}  \revision{either H$\alpha$ or [O III]} over $1 \lesssim z \lesssim 8$. 
\footnote{ \revision{The approximate wavelength coverage of these filters is: F182M (1.722–1.968 $\mu$m), F210M (1.992–2.201 $\mu$m), F250M (2.412–2.595 $\mu$m), F300M (2.831–3.156 $\mu$m), F335M (3.177–3.537 $\mu$m), and F410M (3.865–4.301 $\mu$m).}}
 \revision{We note that the selection for [O III] (H$\alpha$) may not be pure as H$\beta$ ([N II]) can also fall within the filter wavelength range.}
We use F277W, F356W, and F444W to provide the coverage for the broad-band continuum. In addition, F090W, F150W, and F200W are also used as constraints on the galaxy UV size. For [O III] emitters covered by F182M, we also use F210M in addition to F277W for measuring the continuum size, to select galaxies that have strong H$\alpha$ in F277W but not in F210M. Table \ref{tab:filters} summarizes the filter combinations we use and their corresponding redshift coverage for H$\alpha$ ($z_{\rm H\alpha}$) and [O III] 5008\AA\ ($z_{\rm [OIII]}$) lines. 
\rrevision{The redshift ranges $z_{\rm H\alpha}$ and $z_{\rm [OIII]}$ listed in the table include an approximate $\pm10\%$ extension beyond the intrinsic limits defined by the filter bandpasses. This buffer conservatively accounts for uncertainties in photometric redshift estimates (up to 5–10\%) and ensures that potential emission-line features are not excluded due to small mismatches in redshift. We also list the intrinsic redshift ranges separately in the table for reference.} 
$N_{\rm sources}$ is the total number of sources that have photometric redshifts that fall within $z_{\rm H\alpha}$ or $z_{\rm [OIII]}$. We also list the number of outflow candidates that pass our selection process (Section \ref{sec:selection}).

\begin{deluxetable*}{lcccrrr}
\tablenum{1}
\tabletypesize{\small}
\tablecaption{\rrevision{NIRCam Filters Used in This Work}}
\tablehead{
    \colhead{\hspace{0.558cm}Filter combination}\hspace{0.558cm} & 
    \colhead{\hspace{0.558cm}$z_{\rm [O III]}$}\hspace{0.558cm} & 
    \colhead{\hspace{0.558cm}$z_{\rm H\alpha}$}\hspace{0.558cm} & 
    \colhead{\hspace{0.558cm}$z_{\rm [O III],intr.}$}\hspace{0.558cm} & 
    \colhead{\hspace{0.558cm}$z_{\rm H\alpha,intr.}$}\hspace{0.558cm} & 
    \colhead{\hspace{0.558cm}$N_{\rm sources}$}\hspace{0.558cm} & 
    \colhead{\hspace{0.558cm}$N_{\rm candidates}$}\hspace{0.558cm}
}
\decimalcolnumbers
\startdata
F090W/\textbf{F182M}/F277W & 2.19–3.22 & 1.46–2.20 & 2.44–2.93 & 1.62–2.00 & 28231 & 27 \\
F090W/\textbf{F182M}/F210M & 2.19–3.22 & 1.46–2.20 & 2.44–2.93 & 1.62–2.00 & 27499 & 44 \\
F090W/\textbf{F210M}/F277W & 2.68–3.74 & 1.83–2.59 & 2.98–3.40 & 2.03–2.35 & 23349 & 39 \\
F150W/\textbf{F250M}/F356W & 3.43–4.60 & 2.41–3.25 & 3.82–4.18 & 2.67–2.95 & 7813 & 68 \\
F200W/\textbf{F300M}/F356W & 4.19–5.84 & 2.98–4.19 & 4.65–5.30 & 3.31–3.81 & 9663 & 60 \\
F200W/\textbf{F335M}/F444W & 4.81–6.67 & 3.46–4.83 & 5.35–6.06 & 3.84–4.39 & 34822 & 122 \\
F277W/\textbf{F410M}/F444W & 6.05–8.35 & 4.40–6.11 & 6.72–7.59 & 4.89–5.55 & 35816 & 13 \\
\enddata
\tablecomments{\rrevision{Columns: (1) Filter combinations, where the first filter corresponds to the rest-UV, the second is the medium-band filter covering the emission lines (highlighted in bold), and the third traces the continuum; (2) extended redshift range used in candidate selection for [O III] 5008\AA\ emission; (3) extended redshift range used for H$\alpha$ emission; (4) intrinsic redshift coverage for [O III] assuming the exact filter transmission limits; (5) intrinsic redshift coverage for H$\alpha$; (6) number of JADES sources selected using the extended redshift range; (7) number of selected candidates after visual inspection. While the number of sources and candidates are listed together, they could be separated based on the emission line used for detection. Due to overlapping redshift ranges and photometric redshift uncertainty, identifying the number of uniquely detected sources is nontrivial, so we do not attempt to quantify it here.}}
\label{tab:filters}
\end{deluxetable*}

\begin{figure*}[!ht]
    \includegraphics[width=1\textwidth]{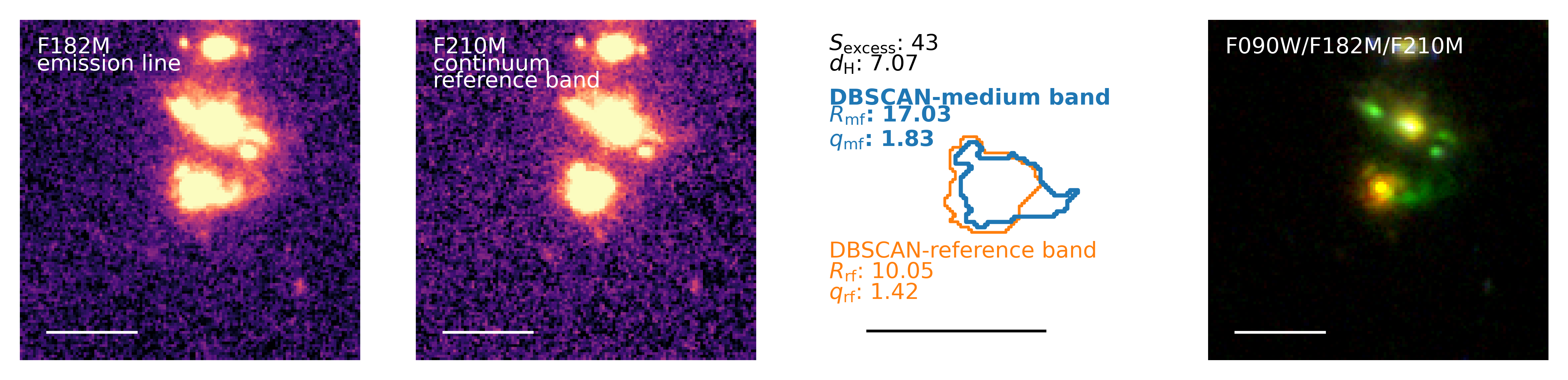}
    \caption{Illustration of the selection procedure. The two panels on the left-hand side show the medium-band and continuum reference band images in log scale. The third panel overlays the medium-band shape (blue) on top of the reference band shape (orange). Shape parameters defined in this paper are labeled in the plot. The right-hand-side panel displays the RGB image (R: F210M, G: F182M, B: F090W) of this galaxy, featuring the potential outflows in green. The scale corresponding to 1\arcsec\ is labeled in each panel. This target has JADES ID$=$209026, and a photometric redshift of $z=2.90$. It is also known as UDF1, a famous ALMA/X-ray/radio AGN \citep{dunlop_deep_2017,xue_chandra_2011,rujopakarn_vla_2016} with a spectroscopic redshift of $z=2.696$ \citep{decarli_alma_2019}.}
    \label{fig:method}
\end{figure*}

\subsection{Selection method} \label{sec:selection}
To retain diffuse and extended emission features as much as possible, we apply a logarithmic stretch on both the medium-band and reference-band \cut{images}  \revision{mosaics for the entire JADES GOODS-S filed}, with a percentile limit of 99.5. \cut{We then cut out the selected galaxies from the re-scaled mosaics into $128\times 128$ pixels$^2$ stamps, which are about $3.84\arcsec\times 3.84\arcsec$ in size. After normalization,}  \revision{The re-scaled mosaics are stored as FITS files, from which we extract $128\times 128$ pixels cutouts (about $3.84\arcsec\times 3.84\arcsec$) for selected galaxies. These cutouts are used as input for the segmentation process.} The edge of each galaxy is detected using the Density-Based Spatial Clustering of Applications with Noise (DBSCAN; \citealp{ester_density-based_1996,schubert_dbscan_2017}) algorithm, which  \revision{operates directly on the 2D flux arrays rather than PNG images commonly used in computer vision applications. To retain the full structure of extended emission, DBSCAN is applied in a 3D parameter space of [x, y, flux], clustering dense regions based on their flux distributions.}

There are two important parameters in DBSCAN: {\tt eps},  \revision{controlling} the maximum distance between two  \revision{pixels (referred to as ``samples'' in DBSCAN terminology)} for one to be considered as in the neighborhood of the other; and {\tt min\_samples}, defining the core size of each cluster. We set {\tt min\_samples=16} to ensure that the saturated galaxy centers can be easily identified as cores (approximately 0.1\arcsec\ in diameter) while ignoring noisy spots. We use {\tt eps=0.09},  \revision{which controls the clustering threshold in the three-dimensional space of [x, y, flux]. The clustering is primarily driven by the flux gradient rather than physical distance alone. This choice results in a typical flux variation of $\sim 0.07$ dex within a cluster.} Our {\tt eps} value is near the optimal choice, as suggested by the ``knee'' in the distance of the nearest neighbor curve as proposed in \cite{rahmah_determination_2016}. We have tested that the clustering results are consistent with the visual impression of multiple team members. 
The typical DBSCAN segmentation edge reaches a surface brightness of approximately 0.02~MJy~sr$^{-1}$ per pixel, corresponding to a 7.7$\sigma$ detection relative to the sky RMS. This corresponds to approximately $m_{\mathrm{AB}} = 28.53$~mag~arcsec$^{-2}$, or $1.5 \times 10^{-18}$~erg~s$^{-1}$~cm$^{-2}$~arcsec$^{-2}$ when integrated over the F300M filter.
In the case of overlapping galaxies, we mask out the companions manually before performing DBSCAN. We caution that although we have attempted to remove large-separation mergers and distinguishable companions, there could still be tidal features that might not be solely caused by feedback.

Since most outflow features are irregular, describing the morphology using conventional quantities, e.g., S\'ersic index \citep{sersic_influence_1963}, asymmetry index \citep{conselice_asymmetry_2000,pawlik_shape_2016}, Gini coefficient \citep{lotz_new_2004}, etc., can be difficult. These metrics typically work well for differentiating disk-like morphologies from merging systems \citep[e.g.,][]{kim_evolution_2021}, while outflows can have very diverse and complicated shapes.  \revision{Although the asymmetry index and Gini coefficient may capture some structural irregularities, their sensitivity to outflows is limited by the relatively low surface brightness of these features compared to the host galaxy. As a result, these traditional metrics may not show significant variations unless the outflow is particularly bright.} Therefore, we define new quantities to describe the morphology of galaxies and quantify the difference between the medium-band and the reference bands.
These quantities are the maximum extent from the centroid of the galaxy, $R$, the excess area in pixel counts, $S_{\rm excess}$, and the maximum axis ratio, $q$. Here, $R$ is given by the maximum distance from the boundary of the DBSCAN shape to the centroid of the galaxy. Since our DBSCAN method is sensitive to the relative instead of absolute change in flux when cutting out the shape, $R$ is stable as long as images in different bands have S/N above our threshold (see below). $S_{\rm excess}$ is calculated by counting the number of pixels that belong only to the cluster in the medium-band. \rrevision{That is to say, we remove pixels that overlap with the broad-band shape from the medium-band pixel count.}
To facilitate direct comparison, we define $S_{\rm rf}$ as the total number of pixels in the reference-band shape.
% \cut{That is to say, we remove pixels that overlap with the broad-band shape from the medium-band pixel count.}  \revision{This is determined by subtracting the pixels that overlap with the broad-band shape from the medium-band pixel count. To facilitate direct comparison, we define $S_{\rm rf}$ as the total number of pixels in the reference-band shape.} 
As for $q$, we first divide the binarized shape from DBSCAN into equal sectors of 15\degr\ around the centroid, and compute the mean distance to the centroid for all pixels in each sector. Then we add up the mean distance from each pair of sectors that are symmetric about the centroid, and $q$ is given by the maximum value divided by the minimum value. For these quantities, we use the subscripts ``mf'' and ``rf'' to represent medium filter and reference filter images, respectively. For reference filter quantities, we take the maximum values from the bluer or redder bands listed in Table \ref{tab:filters}. All quantities are computed without matching the PSF in each band to avoid introducing artifacts in morphology measurements, increasing the apparent galaxy size in \cut{medium band}  \revision{the medium-band image}, and decreasing the signal-to-noise ratio.

We aim to select galaxies that have much greater extent, very different axis ratio, or many excess pixels in the medium band compared to the reference bands.  \revision{A galaxy is selected if it meets at least one of the following criteria:}
\begin{itemize}
    \item $R_{\rm mf} > 1.5R_{\rm rf}$, or
    \item $|q_{\rm mf} - q_{\rm rf}| > 0.3$, or
    \item $S_{\rm excess} > 0.3S_{\rm rf}$, where $S_{\rm rf}$ is the pixel count in the reference bands.
\end{itemize}
We also require detection $S/N>10$ for all bands in the filter combination based on the median flux of the brightest 10 pixels relative to the RMS noise. 
Finally, we perform a visual inspection to exclude suspicious selections.  \revision{Out of the total objects initially selected based on the quantitative criteria, 1066 unique candidates were identified, of which 326 were confirmed after visual inspection, while 740 were discarded due to contamination, artifacts, or ambiguous morphology.}
% Finally, we perform a visual inspection to exclude suspicious selections. 
Figure \ref{fig:method} illustrates our clustering and selection procedure. In the figure, we include an additional quantity, the Hausdorff distance ($d_{\rm H}$; \citealp[see][p.~177]{rockafellar_variational_2004}). The Hausdorff distance between two sets of points $A$ and $B$ is defined as:
\begin{equation}
d_{\rm H}(A, B) = \max \left\{ \sup_{a \in A} \inf_{b \in B} d(a, b), \sup_{b \in B} \inf_{a \in A} d(a, b) \right\},
\end{equation}
where $d(a, b)$ denotes the Euclidean distance between points $a$ and $b$.
It measures how far the two shapes are from being isometric. 
We do not use $d_{\rm H}$ in the sample selection phase. Nevertheless, it is an important parameter indicating the geometry of the extended emission feature (see Section \ref{sec:agn}).

\subsection{Spectroscopic observations} 
To probe the kinematics of the selected candidates, we also include NIRSpec MSA observations from JADES DR3 \citep{deugenio_jades_2024} and SMILES (\citealp{alberts_smiles_2024}; Y.~Zhu et al. in preparation), where available. Both data sets cover the GOODS-S field and provide medium resolution ($R \sim 1000$) grating spectra. We find that 5 of our candidates have G140M/F100LP and G235M/F170LP spectra covering $0.97 \mu{\rm m} < \lambda < 3.07 \mu{\rm m}$ from SMILES, and 8 have G140M/F070LP, G235M/F170LP, and G395M/F290LP spectra covering $0.70 \mu{\rm m} < \lambda < 5.10 \mu{\rm m}$ from JADES. The objects with spectra and the potential outflow features are discussed in Section \ref{sec:spectra}.

\section{Overview of the Selection Results}\label{sec:results}

\begin{figure*}[!ht]
    \centering
    \includegraphics[width=1\textwidth]{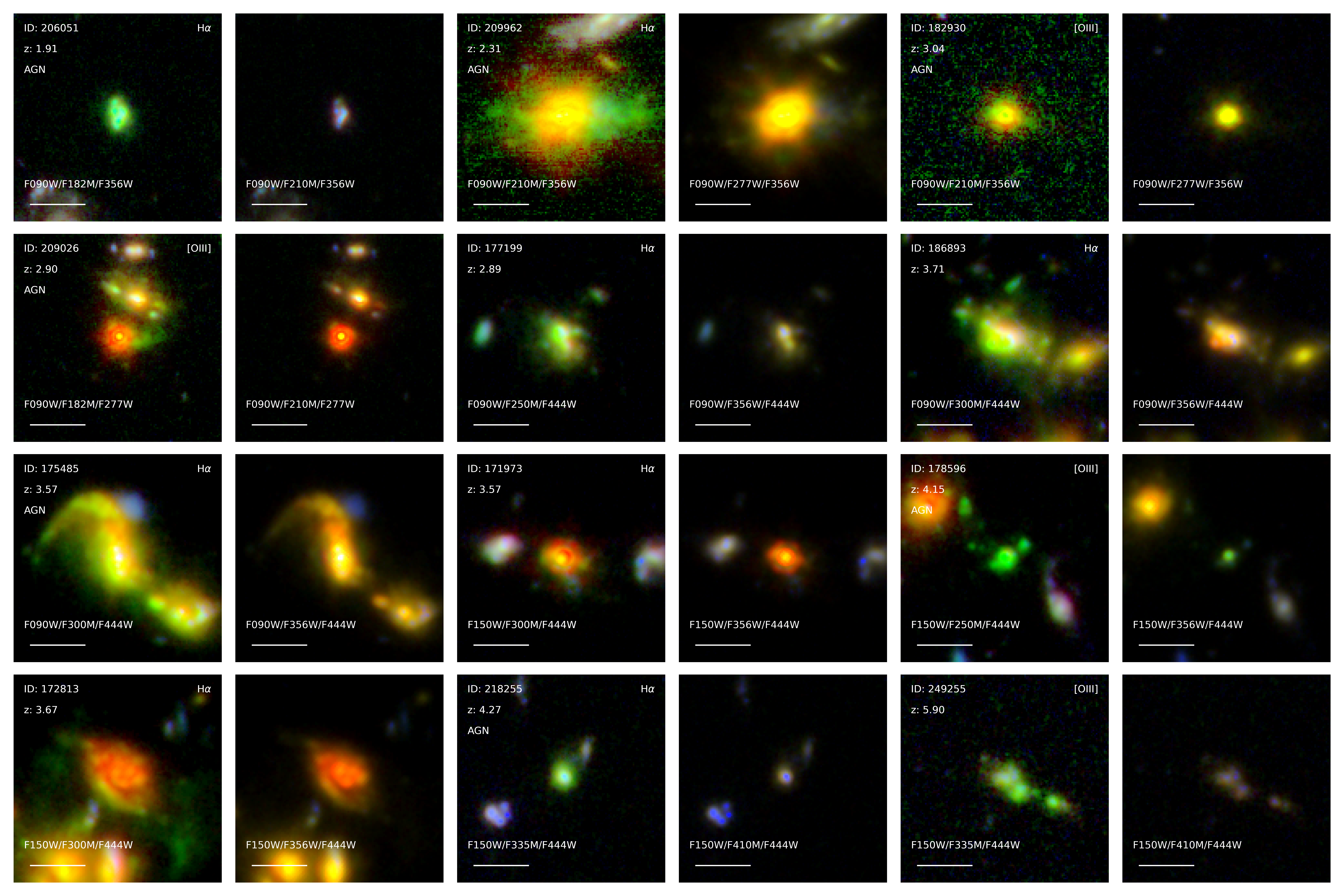}
    \caption{Examples of selected outflow / extended emission line candidates. The first, third, and fifth columns show galaxy images with the medium-band listed in Table \ref{tab:filters} being the green channel, and the corresponding continuum RGB images are shown in the second, fourth, and sixth columns, respectively. The comparison highlights emission extending beyond the stellar continuum in our selected galaxies. The white bar denotes the scale of 1\arcsec. We also label the \textit{photometric} redshift,  \revision{the emission line used for identification}, and the presence of known AGN in the figure. We note that some of the galaxies shown here have already been observed in the literature. For example, ID 175485 (MIRI AGN, $z_{\rm phot}=3.57$) has NIRSpec/IFS observations (GA-NIFS ID: GS-5001, $z_{\rm spec}=3.47$) and similar outflows are identified in the central galaxy \citep[][although they do not find a clear indication of AGN]{lamperti_ga-nifs_2024}. ID 209962 ($z_{\rm phot}=2.31$) is also known as K20-ID5, and its kinematics and outflow properties have been studied with KMOS and SINFONI data ($z_{\rm spec}=2.224$; \citealp{forster_schreiber_sinszc-sinf_2014,genzel_evidence_2014,loiacono_multiwavelength_2019,scholtz_kashz_2020,davies_nuclear_2020}).  \revision{The color and contrast in this plot have been adjusted to better visualize the faint and extended emission lines.}}
    \label{fig:example}
\end{figure*} 

\begin{figure*}[!ht]
    \gridline{
    \fig{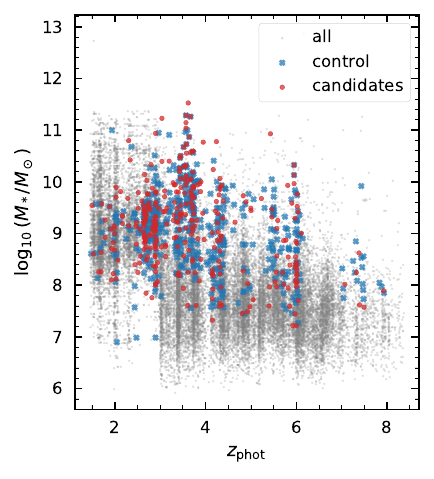}{0.33\textwidth}{(a)}
    \fig{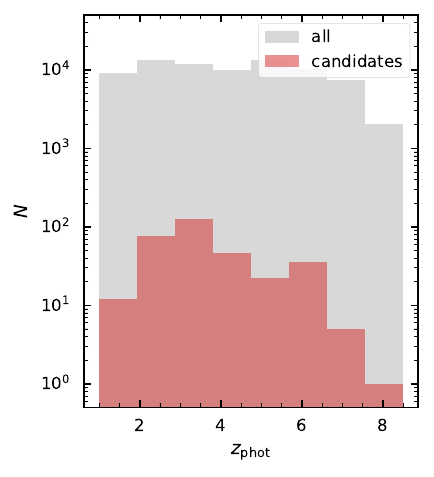}{0.33\textwidth}{(b)}
    \fig{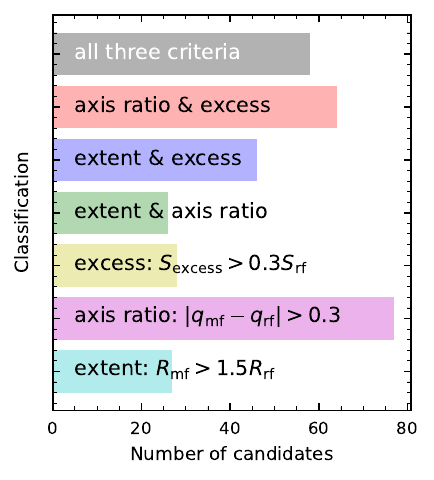}{0.33\textwidth}{(c)}}
    \caption{Overview statistics of the parent sample and our candidates. (a) Scatter plot of the stellar mass and photometric redshift for our candidates (red dots) and the control sample (blue crosses). We also show the parent sample for reference \revision{, which consists of all JADES GOODS-S sources with medium-band coverage (Table \ref{tab:filters}). The SED properties for the parent sample shown in this figure are taken from ASTRODEEP \citep{merlin_astrodeep-gs43_2021} for $z < 3$ galaxies and from \citet{simmonds_ionising_2024} for $z > 3$ galaxies. The apparent break in stellar masses at $z \sim 3$ reflects this transition in the SED property derivation for the parent sample.} (b) Redshift distribution of the parent sample and our candidates. The overall selection rate is approximately $\sim 1\%$.  \revision{(c) Number counts of the selected candidates in each category. Each bar represents galaxies that fulfill \textbf{only} the corresponding selection criterion or combination of criteria, and no bar is included within any other. For example, galaxies in the yellow bar are selected \textbf{only} by the excess criterion $S_{\rm excess} > 0.3 S_{\rm rf}$ and not by extent or axis ratio thresholds.}}
    \label{fig:overview}
\end{figure*}

\begin{figure*}[!ht]
    \centering \includegraphics[width=1.0\textwidth]{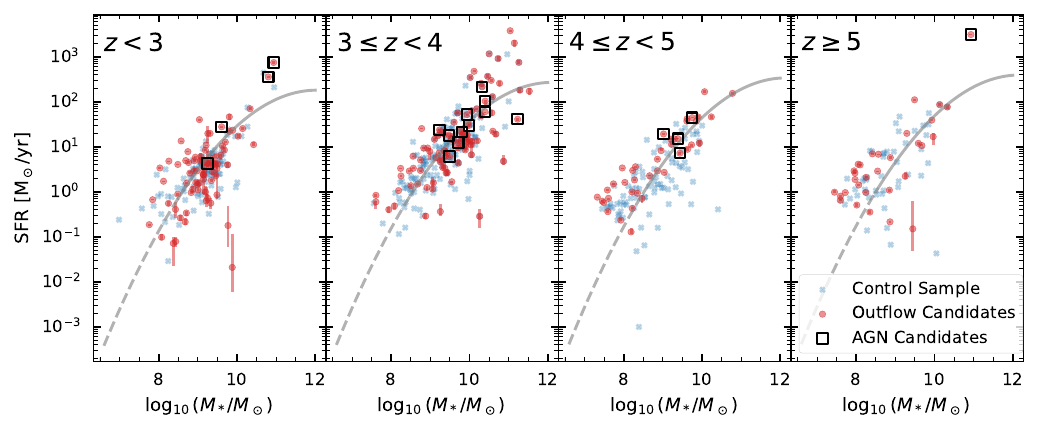}
    \caption{The relationship between star formation rate (averaged over 30 Myr) and stellar mass for galaxies in each redshift bin. Candidates and control sample are shown in red and blue, respectively. Candidates with AGN are marked with black boxes  \revision{and are just for reference as their SED-derived properties might be biased}. For reference, the gray curve plots the star-forming main sequence from \citet{popesso_main_2023}, with the dashed line being our extrapolation. Overall, our candidates are actively forming stars.
    }
    \label{fig:SFR_Mstar}
\end{figure*}

\begin{deluxetable*}{cccccccccccccccc}
\tablenum{2}
\tabletypesize{\footnotesize}
\tablecaption{Properties of Medium-Band Selected Candidates}
\tablehead{
    \colhead{ID} & 
    \colhead{RA} & 
    \colhead{Dec} & 
    \colhead{$z_{\rm phot}$} & 
    \colhead{ \revision{Line}} & 
    \colhead{AGN} & 
    \colhead{Med. filter} & 
    \colhead{Ref. filter} & 
    \colhead{$R_{\rm mf}$} & 
    \colhead{$R_{\rm rf}$} & 
    \colhead{$q_{\rm mf}$} & 
    \colhead{$q_{\rm rf}$} & 
    \colhead{$S_{\rm mf}$} & 
    \colhead{$S_{\rm rf}$} & 
    \colhead{$S_{\rm excess}$} & 
    \colhead{$d_{\rm H}$}
}
\decimalcolnumbers
\startdata
208077 & 53.133284 & -27.779891 & 3.55 & Halpha & False & F335M & F444W & 21.10 & 19.42 & 2.46 & 3.32 & 332 & 163 & 169 & 3.16 \\
208828 & 53.120050 & -27.777855 & 2.69 & [OIII] & False & F182M & F277W & 9.49 & 7.28 & 2.04 & 1.73 & 82 & 78 & 16 & 4.00 \\
209026 & 53.183472 & -27.776656 & 2.90 & [OIII] & True & F182M & F210M & 17.03 & 10.05 & 1.83 & 1.42 & 200 & 218 & 43 & 7.07 \\
209960 & 53.130468 & -27.773418 & 2.17 & Halpha & False & F210M & F277W & 7.07 & 5.39 & 2.21 & 1.59 & 73 & 60 & 17 & 2.24 \\
209962 & 53.131149 & -27.773190 & 2.31 & Halpha & True & F210M & F277W & 56.08 & 36.06 & 2.00 & 1.50 & 2610 & 1699 & 936 & 22.09 \\
210132 & 53.190617 & -27.773266 & 3.38 & [OIII] & False & F210M & F277W & 17.03 & 12.37 & 2.81 & 3.39 & 137 & 46 & 91 & 5.39 \\
210963 & 53.176566 & -27.771131 & 5.70 & [OIII] & False & F335M & F444W & 12.21 & 12.04 & 1.94 & 2.68 & 164 & 139 & 28 & 3.61 \\
211355 & 53.135714 & -27.768807 & 3.39 & [OIII] & True & F210M & F277W & 17.09 & 9.85 & 2.06 & 1.54 & 206 & 145 & 67 & 7.28 \\
211831 & 53.175615 & -27.768818 & 3.58 & Halpha & False & F335M & F444W & 12.21 & 9.85 & 1.62 & 1.78 & 249 & 149 & 100 & 3.16 \\
212023 & 53.195260 & -27.767929 & 2.90 & [OIII] & False & F210M & F277W & 16.64 & 16.12 & 1.56 & 2.18 & 266 & 184 & 97 & 7.00 \\
212228 & 53.203629 & -27.767469 & 4.22 & Halpha & True & F335M & F444W & 16.76 & 14.87 & 2.15 & 2.81 & 341 & 184 & 157 & 3.61
\enddata
\tablecomments{
Columns:
(1) ID of the galaxy in JADES GOODS-South catalog v0.9.3;  
(2) \& (3) Coordinates in J2000;  
(4) Photometric redshift;  
(5)  \revision{Primary detection line (H$\alpha$ or [O III]); if both are detected, the most prominent one is listed;}
(6) Whether the galaxy hosts a known AGN, as identified in \citet{lyu_active_2024}, \citet{matthee_little_2024}, and \citet{sun_no_2025};  
(7) Medium band used;  
(8) Adjacent reference band for continuum;  
(9) Maximum extent from the centroid in the medium band;  
(10) Maximum extent from the centroid in the reference band;  
(11) Maximum axis ratio in the medium band;  
(12) Maximum axis ratio in the reference band;  
(13) Area of the DBSCAN clustering in the medium band;  
(14) Area of the DBSCAN clustering in the reference band;  
(15) Excess area in number of pixels that are only present in the medium-band shape;  
(16) Hausdorff distance between the medium-band shape and the reference-band shape.  
\\
(This table is available in its entirety in machine-readable form.)
}
\label{tab:candidates}
\end{deluxetable*}

In total, we selected 326 galaxies with potential outflows or extended emission-line features.  
\rrevision{Among them, 172 are selected based on H$\alpha$, 134 based on [O III], and 20 are selected by both lines. While a single filter cannot cover both H$\alpha$ and [O III] simultaneously, the overlap in redshift coverage from different medium-band combinations (combined with the $\sim$10\% photometric redshift allowance) can result in a small number of galaxies being identified in both categories.}
Figure \ref{fig:example} showcases 12 galaxies in our sample. The images with the medium-band as the green channel highlight the feature of interest in green, compared to their corresponding RGB images based on the stellar continuum. The emission features that extend beyond the continuum show a great diversity in their scale, structure, and brightness. Also, the host galaxies can be very different from each other in their morphologies. \cut{We list their properties, mainly the morphological parameters used in this work, in Table}  \revision{Most galaxies are selected based on only one emission line ([O III] or H$\alpha$) due to differences in intrinsic emission-line strength relative to the continuum and the larger PSF at longer wavelengths, which can dilute the contrast between the medium-band and adjacent continuum-band fluxes. We list their properties, including the primary detection line and morphological parameters, in Table \ref{tab:candidates}.}

 \revision{Table \ref{tab:candidates} also lists whether a candidate galaxy has AGN. In this work, we use the AGN catalog in \citet{lyu_active_2024} based on MIRI measurements from the SMILES survey \citep{alberts_smiles_2024, rieke_smiles_2024}  \revision{targeting the GOODS-S/HUDF field}. We also use the broad emission line identified AGN sample in \citet{sun_no_2025} and \citet{matthee_little_2024} based on NIRCam/grism spectra in the FRESCO \citep{oesch_jwst_2023} survey.  \revision{These selection methods are complementary, and the only broad-line AGN in our candidate sample is included in all these catalogs.} \footnote{The galaxy with JADES ID 333792 is identified as an AGN in \citet{matthee_little_2024} (GOODS-S-13971) and is included in our outflow candidates.}}

To derive the physical properties of the galaxies, we use {\tt Prospector} \citep{johnson_stellar_2021} to fit the spectral energy distribution (SED) based on the multi-band Kron convolved photometry from JADES \citep{eisenstein_jades_2023,bunker_jades_2024} and the legacy Hubble Ultra Deep Field (HUDF; \citealp{beckwith_hubble_2006}; also see CANDELS: \citealp{grogin_candels_2011,koekemoer_candels_2011}) data following \citet{simmonds_low-mass_2024}. 
\footnote{When performing the SED fitting, we assume that the flux is dominated by the stellar population emission without decomposing AGN, if any.}
The filters we use include the HST ACS bands: F435W, F606W, F775W, F814W, and F850LP; the HST WFC3/IR bands: F105W, F125W, F140W, and F160W; and the JWST NIRCam bands: F070W, F090W, F115W, F150W, F162M, F182M, F200W, F210M, F250M, F277W, F300M, F335M, F356W, F410M, F430M, F444W, F460M, and F480M. We adopt the assumptions and priors in \citet{ji_jades_2023} when fitting the SEDs. Briefly, we use a non-parametric star formation history (SFH) described by \citet{leja_how_2019}, modeled as nine SFR bins controlled by the continuity prior. We use the Kroupa \citep{kroupa_variation_2001} stellar initial mass function (IMF). The continuum and emission properties are generated using the Flexible Stellar Population Synthesis (FSPS) code \citep{byler_nebular_2017} based on Cloudy models \citep{ferland_2013_2013} and the MILES stellar library \citep{vazdekis_evolutionary_2015}. Following \citet{tacchella_fast_2022}, we treat dust attenuation from young stars (age $<$ 10 Myr) and nebular emission lines, and old stars (age $>$ 10 Myr) differently \citep{charlot_simple_2000}. Here, the dust attenuation law for the old stellar population is parameterized as in \citet{noll_analysis_2009}, with the UV dust bump tied to the dust index based on the results in \citet{kriek_dust_2013}. As for the IGM transmission, we use the \citet{madau_radiative_1995} model. In addition to SED fitting performed in this work, we also include galaxy properties presented in \citet{simmonds_ionising_2024} for $z>3$ JADES galaxies for reference (Appendix \ref{app:simmonds}), where additional ionizing properties are measured.  \revision{We caution that our SED fitting does not include a reliable AGN model; therefore, we exclude AGNs from all analyses involving SED-fitting-based galaxy properties.}

To find out whether the selected candidates occupy a unique position in the galaxy property space, we build a control sample for comparison in the sections below. For each galaxy in our candidate sample, we select 3 non-candidate galaxies from the parent sample, with the closest reference band flux and photometric redshift. We allow a difference in flux of 15\% and $\pm 0.15$ in photometric redshift.  \revision{In 96\% of cases, three matched control galaxies are found. However, 1\% of candidates have only one or two matches, and 3\% have none within the selection criteria.}

Figure \ref{fig:overview} provides an overview of our selection results. Our sample spans a large redshift range of \cut{$1.5 \lesssim z \lesssim 6.5$}  \revision{$1.4 \lesssim z \lesssim 8.4$}. The stellar mass also covers the distribution of the parent sample over $7 \lesssim \log_{10}(M_*/M_\sun) \lesssim 11.5$.  \revision{The parent sample includes all JADES GOODS-S sources that have medium-band coverage in Table \ref{tab:filters}, with SED properties derived from ASTRODEEP \citep{merlin_astrodeep-gs43_2021} for $z < 3$ galaxies and from \citet{simmonds_ionising_2024} for $z > 3$ galaxies.
While the underlying parent sample is relatively homogeneous for $z \gtrsim 3$, we find that the stellar masses of our selected candidates (and consequently the control sample) tend to decrease with redshift. At $3 < z < 4$, only a small fraction of galaxies have $\log_{10}(M_*/M_\sun) < 8.5$, whereas at $5 < z < 6$, this fraction rises to nearly half, and at $z > 6$, almost all candidates fall below this threshold. Additionally, most of the highest-mass galaxies at $z \sim 3.5$ are selected as candidates, a trend less apparent at lower redshifts.
} 
As shown in Figure 3(b), the selection rate is about 1\% \cut{and no strong evolution with redshift is observed, except for a peak in the population near cosmic noon.}  \revision{, with a peak near $z \sim 3$. However, it is unclear whether this reflects a true redshift evolution or is influenced by selection effects.}
Figure \ref{fig:overview}(c) shows the number of candidates selected by different criteria. About 20\% of the candidates are selected because of their significantly different maximum axis ratios between the medium-band and the reference bands. The threshold on excess pixels alone selects $\sim 10\%$ of the candidates, while the maximum extent condition selects about another 10 percent. Galaxies in the rest of the sample fulfill multiple criteria.

We further explore the overall star-forming properties of our candidates using the SED fitting results. We divide the sample into four redshift bins for $z<3$, $3 \leq z < 4$, $4 \leq z < 5$, and $z \geq 5$, and plot the star formation rate (SFR) versus stellar mass ($M_*$) in Figure \ref{fig:SFR_Mstar}. We observe a roughly positive correlation between SFR and $M_*$, with no significant evolution between redshift bins. As an important reference, we plot the star-forming main sequence proposed by \citet{popesso_main_2023}, with our extrapolation below $M_* \lesssim 10^8 M_\sun$, as the low-mass end behavior is still not well determined. 
In general, our candidates are located on or above the star-forming main sequence in all redshift bins, showing active star-forming activities \citep[also see][]{rinaldi_galaxy_2022,rinaldi_midis_2024}.  \revision{Specifically, the fraction of outflow candidates above the main sequence by more than 0.2 dex is approximately 29\% at $z < 3$, 36\% at $3 \leq z < 4$, 60\% at $4 \leq z < 5$, and 65\% at $z \geq 5$. The fraction below the main sequence by more than 0.2 dex is 46\%, 35\%, 15\%, and 26\% in these respective redshift bins, while the remaining candidates lie within $\pm 0.2$ dex of the main sequence.}
While we see some massive candidates ($M_* \gtrsim 10^{10} M_\sun$) with high SFRs, intriguingly, many of the candidates are low in  stellar mass, with $M_* \lesssim 10^8 M_\sun$. These low-mass galaxies, however, are very active in forming new stars. \cut{We do not see a significant difference between candidates with or without AGN in the SFR-$M_*$ diagram, except that candidates with identified AGN are more massive.} A detailed comparison between the candidates and the control sample is presented in the following section.

\section{Discussion}\label{sec:discussion}

We aim to understand the mechanisms driving the observed medium-band morphologies that may indicate potential strong outflows or extended emission lines. In this section, we perform detailed investigations on the galaxy properties between our candidates and the control sample, the correlation between the extent of the DBSCAN shape and galaxy parameters, and differences between AGN vs non-AGN. We also \cut{present}  \revision{discuss} spectroscopic observations for a subset of the sample.

\subsection{Candidates \textit{vs.} Control Sample}

\begin{figure*}[!ht]
    \centering
    \includegraphics[width=1\textwidth]{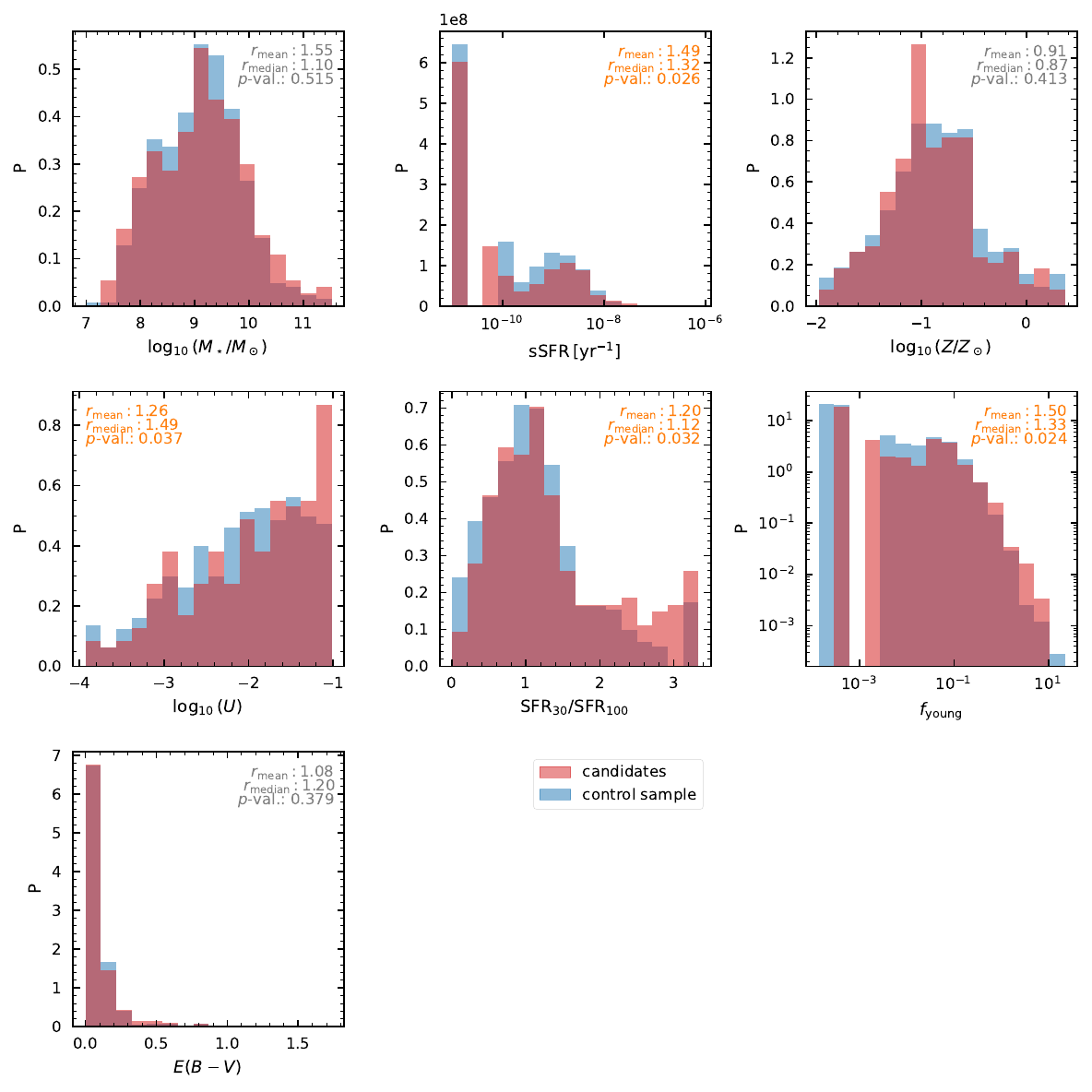}
    \caption{Comparison of galaxy properties from SED fitting between our  \revision{non-AGN} outflow candidates and the control sample. In each panel, we plot the histogram of the \cut{probability density}  \revision{density distribution} of the two samples, with red and blue showing the candidates and the control sample, respectively. If there is a significant difference between the two, i.e., $p$-value for the K-S test is less than 0.05, we \cut{label}  \revision{highlight} the ratio of mean value, ratio of median value, and the $p$-value in the corresponding panels.}
    \label{fig:candidates_vs_control}
\end{figure*}

 \revision{Since our SED model does not reliably account for AGN activity, we have excluded all AGNs from the analysis in this subsection.} 
Figure \ref{fig:candidates_vs_control} compares various galaxy properties between our selected candidates and a control sample. As described in Section \ref{sec:data}, the control sample is selected to have similar redshift and continuum flux, ensuring that the stellar mass distributions are comparable across both samples. Despite the similarities in stellar mass, the histograms of the \cut{probability density}  \revision{density distribution} for several key properties reveal significant differences between the candidates and the control sample.

The distribution of $M_*$ indicates that both the candidates and control sample are matched well in terms of stellar mass, as expected. The metallicity distributions between the candidates and the control sample show considerable overlap. Nevertheless, the specific star formation rate (sSFR) shows a notable difference. The mean and median sSFR values are higher for the candidates, and we calculate the ratios ($r$) of sSFR between the candidate sample and the control sample value to be $r_{\rm{mean}} = 1.49$ and $r_{\rm{median}} = 1.32$. This suggests that our selected candidates are forming stars at a  higher rate relative to their stellar mass compared to the control sample. 
We note that stellar-feedback-driven outflows have been shown to correlate with SFR properties at cosmic noon \citep[e.g.,][]{forster_schreiber_kmos3d_2019}. % HU
The K-S test $p$-value of 0.026 further confirms this difference, indicating a statistically significant ($p<0.05$) enhancement in star formation activity among the candidates.

A significant difference is observed in the ionization parameter ($U$) distributions. The candidates exhibit higher mean and median ionization parameters, with a ratio $r_{\rm{mean}} = 1.26$ and $r_{\rm{median}} = 1.49$, and a $p$-value of 0.037. This indicates more highly ionized environments in our candidate galaxies. 

Appendix \ref{app:simmonds} also presents the measurements of ionizing properties from \citet{simmonds_ionising_2024},
\footnote{ \revision{Our SED fitting method does not directly measure the ionizing photon production efficiency and production rate. Therefore, here we use the measurements from \citet{simmonds_ionising_2024}.}}
 \revision{and similar correlations are found}. In Figure \ref{fig:candidates_vs_controlc}, the ionizing photon production efficiency ($\xi_{\rm{ion}}$)
\footnote{ \revision{The ionizing photon production efficiency measured here assumes an ionizing escape fraction of zero.}}
shows a considerable difference, with candidates having higher $\xi_{\rm{ion}}$ values ($r_{\rm{mean}} = 1.17$, $r_{\rm{median}} = 1.15$, $p$-value = 0.065). These elevated ionization parameters and $\xi_{\rm{ion}}$ values suggest that candidates are capable of producing and sustaining more ionized gas, likely driven by intense star formation and possible AGN activity \citep[][]{rinaldi_midis_2024}. 
% comment 403
 \revision{The systematically higher $\xi_{\rm ion}$ values in our candidates are in agreement with trends observed in MOSDEF galaxies at $z\sim2$ \citep{shivaei_mosdef_2018}, where galaxies with stronger ionizing continua tend to be associated with younger stellar populations and more active star formation.}
The ionizing photon production rate ($\dot{n}_{\rm{ion}}$) further underscores the differences between the two samples. Candidates have significantly higher $\dot{n}_{\rm{ion}}$ values ($r_{\rm{mean}} = 1.66$, $r_{\rm{median}} = 1.50$, $p$-value = 0.031), which aligns with the findings of higher sSFR and $\xi_{\rm{ion}}$. This suggests that candidates are not only forming stars more rapidly but also contributing more to the ionizing photon budget of the universe, impacting their surrounding environments and possibly driving the extended emission lines and potential strong outflows \citep[e.g.,][]{bugiani_agn_2024,rinaldi_midis_2024}.

The burstiness of star formation and the fraction of young stellar mass also exhibit clear distinctions. Here, we calculate burstiness as $\rm SFR_{30}/SFR_{100}$, where $\rm SFR_{30}$ is the averaged star formation rate in the most recent SFH bin (recent 30 Myr) and $\rm SFR_{100}$ is the averaged star formation rate in the recent 100 Myr \citep[see e.g.,][and references therein]{simmonds_low-mass_2024}. The fraction of young stellar mass ($f_{\rm young}$) is the fraction of the stellar mass formed in the most recent bin (30 Myr) in the SFH out of the total stellar mass. Candidates are shown to be $\sim 20\%$ more bursty in their star formation activities, with $r_{\rm{mean}} = 1.20$ and $r_{\rm{median}} = 1.12$, and a $p$-value of 0.032. This is supported by a higher $f_{\rm young}$ ($r_{\rm{mean}} = 1.50$, $r_{\rm{median}} = 1.33$, and a $p$-value of 0.024) in the candidates compared to the control sample. We note that if we compute the burstiness based on $\rm SFR_{10}/SFR_{100}$, the candidates can be $\sim30\%$ more bursty than the control sample (see Figure \ref{fig:candidates_vs_controlc}). These findings suggest that the candidates are experiencing more recent and intense episodes of star formation activities, aligning with the higher production rate of ionizing photons discussed above \citep[e.g.,][]{faisst_recent_2019,atek_star_2022}.

Interestingly, the distributions of dust attenuation ($E(B-V)$) do not show a significant difference between the candidates and the control sample, as the $p$-value significantly exceeds 0.05. This indicates that while candidates are more active in star formation and have more ionized gas, their overall dust content and attenuation properties are comparable to the control sample. This could imply that the differences observed in star formation and ionization parameters are intrinsic to the stellar populations and gas properties rather than being significantly influenced by dust. In summary, our selected candidates exhibit significantly higher specific star formation rates, ionization parameters, ionizing photon production efficiencies, burstiness, and young stellar mass fractions compared to the control sample.

\subsection{Physical properties and galaxy morphology}
\begin{figure*}[!ht]
    \centering
    \includegraphics[width=1\textwidth]{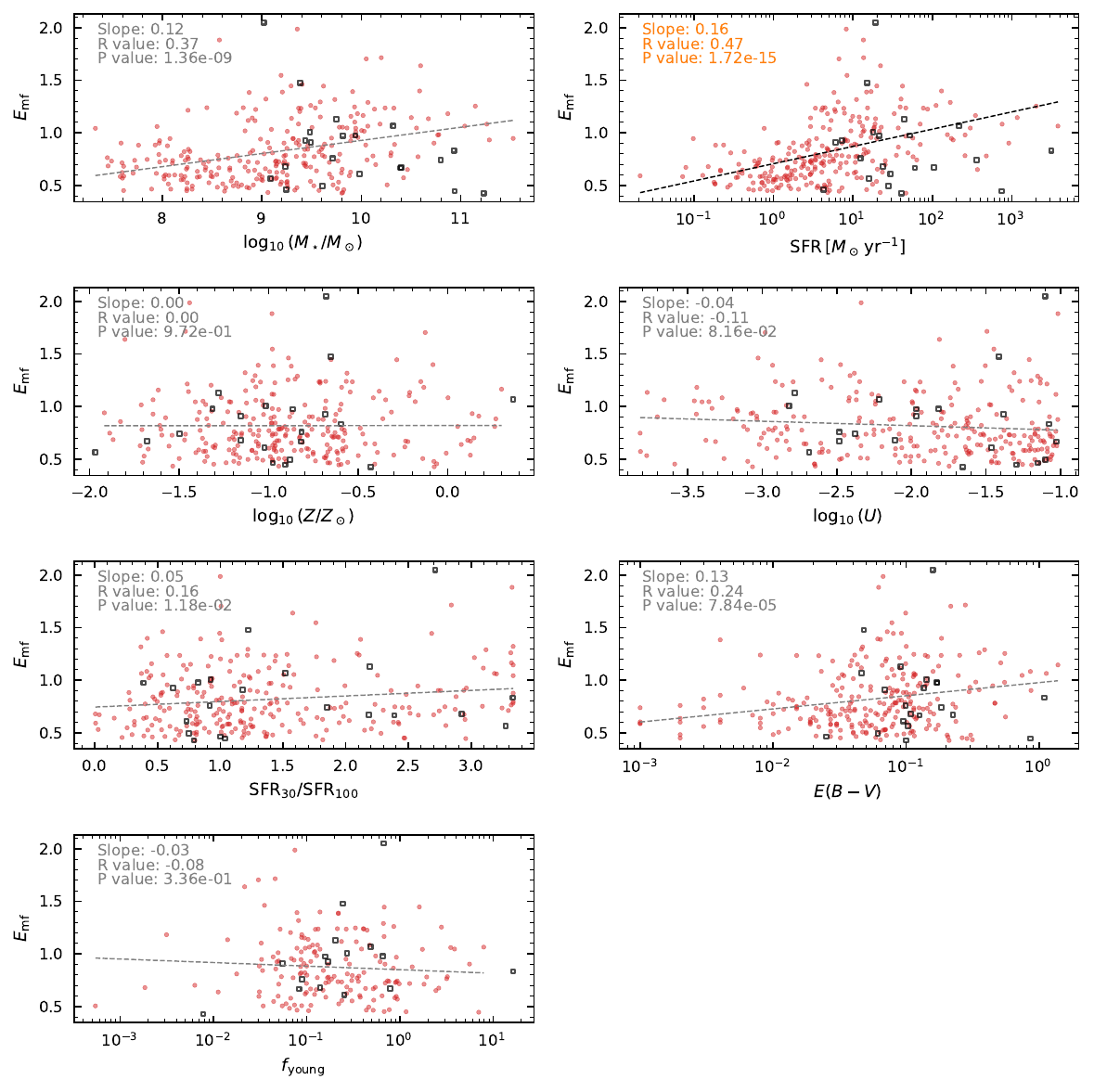}
    \caption{Relation between the extent of potential outflows or extended emission lines with galaxy properties based on SED fitting. We perform a linear fit between $E_{\rm mf} = \sqrt{S_{\rm excess} / S_{\rm rf}}$ and the common logarithm of galactic parameters. Black boxes mark objects hosting known AGN  \revision{just for reference and are not included in the fitting}. The slope, correlation coefficient ($R$), and $p$-value of the linear fits are labeled in each panel. We do not see strong correlations ($R>0.4$) between the extent of extended emission and galaxy properties, except for SFR (and $\dot{n}_{\rm ion}$ in Figure \ref{fig:ext_vs_parametersc}).}
    \label{fig:ext_vs_parameters}
\end{figure*}

 \revision{Figure \ref{fig:ext_vs_parameters} plots the relationship between the extent of potential outflows or extended emission lines, quantified as $E_{\rm mf} = \sqrt{S_{\rm excess}/S_{\rm rf}}$, and various galaxy properties derived from SED fitting. Each panel displays the linear fit between $E_{\rm mf}$ and the common logarithm of a specific galactic parameter, with the slope, correlation coefficient ($R$), and $p$-value of the fit annotated. The p-value is derived from a significance test of the Pearson correlation coefficient, assessing the likelihood of obtaining the observed correlation by chance.}  \revision{Since SED-derived galaxy parameters can be highly biased for AGN hosts, we include them in the plot just for reference but do not use them for fitting or further analysis.}

The star formation rate shows the strongest correlation with $E_{\rm mf}$. The SFR has an $R$ value of 0.47 and a highly significant $p$-value of $1.72 \times 10^{-15}$, suggesting that galaxies with higher SFRs are more likely to have extensive outflows or extended emission lines. As star formation is a key driver of outflows, this result is consistent with our expectations  \revision{ \citep[e.g.,][]{veilleux_galactic_2005,heckman_cos-burst_2017}}. The ionizing photon production rate, which is closely associated with star formation activities, also shows a strong correlation with $E_{\rm mf}$, with $R = 0.47$ and a $p$-value of $1.87 \times 10^{-11}$ (see Figure \ref{fig:ext_vs_parametersc}). This further highlights the importance of active star formation and ionizing photon production in driving the extent of ionized regions and outflow features.

Other galaxy properties, such as stellar mass, show more moderate correlations with $E_{\rm mf}$. $M_\star$ has an $R$ value of 0.37, indicating some association but not as strong as SFR. This weaker correlation is likely a second-order effect, given the known relation between stellar mass and SFR through the star-forming main sequence. Similarly, dust attenuation shows a weaker correlation ($R = 0.24$), which is also expected due to the dust-stellar mass relation.

We find weaker correlations with metallicity ($R = 0.00$) and the ionization parameter ($R = -0.11$). Burstiness and young stellar mass fraction exhibit similarly weak correlations, with $R = 0.16$ ($p$-value of 0.01) and $R = -0.08$ ($p$-value of 0.34), respectively. These properties, while statistically significant in some cases, do not appear to be strong drivers of the extent of outflows or extended emission lines. The weaker correlations between $M_\star$ (or SFR) and other quantities like  \revision{$\log U$} and burstiness suggest that these are likely secondary effects related to the overall star formation activity.

\subsection{AGN \textit{vs.} non-AGN}\label{sec:agn}

\begin{figure*}[!ht]
    \centering
    \includegraphics[width=1.0\textwidth]{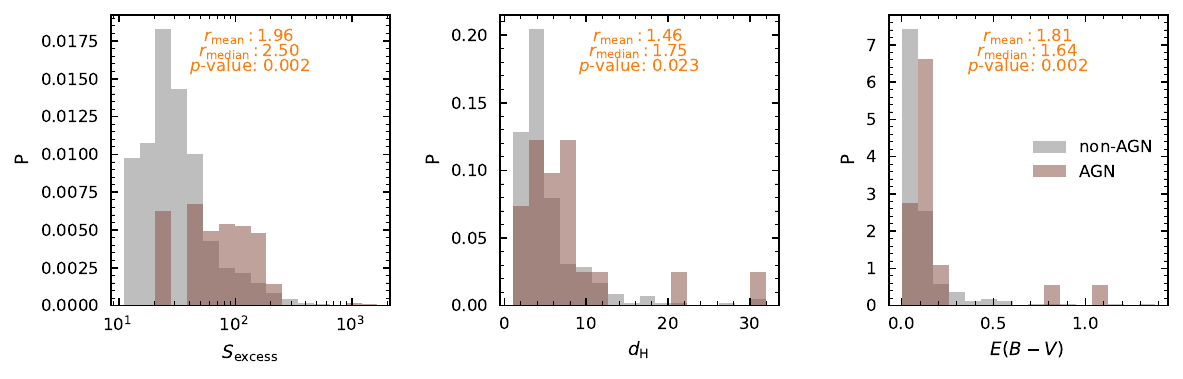}
    \caption{Among candidates with potential outflows, comparison of galaxy properties between AGN and non-AGN. Panels from left to right show the \cut{probability} density distribution for excess area in the medium-band, the Hausdorff distance between the medium-band shape and the reference band shape, and the dust attenuation ($E(B-V)$). Brown and gray colors represent the candidates hosting AGN and non-AGN candidates, respectively. We label the ratio of mean, median, and K-S test $p$-values in each panel.}
    \label{fig:agn_vs_nonagn}
\end{figure*}

Figure \ref{fig:agn_vs_nonagn} compares the properties of candidate galaxies with and without known AGN  \revision{(see Section \ref{sec:results})} to investigate the potential origins of outflows and the morphology of extended emission. The panels show the \cut{probability} density distribution for the excess area in the medium-band ($S_{\rm excess}$), the Hausdorff distance between the medium-band shape and the reference band shape, and the dust attenuation parameter ($E(B-V)$). Red and blue colors represent the AGN candidates and the non-AGN control sample, respectively. The ratios of mean, median, and K-S test $p$-values are labeled in each panel.

The excess area ($S_{\rm excess}$) shows a significant difference between AGN and non-AGN candidates. AGN candidates tend to have larger $S_{\rm excess}$ values, with a mean ratio of 1.96 and a median ratio of 2.50, supported by a $p$-value of 0.002. This indicates that galaxies hosting AGN are more likely to exhibit extensive outflows or extended emission lines. 
% comment 403:
 \revision{This trend is consistent with findings from KMOS3D \citep{forster_schreiber_kmos3d_2019}, where AGN-driven outflows are found to be more prevalent and stronger in massive galaxies.}
This finding is  \revision{also} consistent with the understanding that AGN-driven outflows can significantly impact the surrounding interstellar medium, often producing more prominent and widespread ionized regions \citep[e.g.,][]{king_powerful_2015, davies_jwst_2024}.

The Hausdorff distance between the medium-band shape and the reference band shape also shows a significant difference, with AGN candidates having higher values (mean ratio of 1.46 and median ratio of 1.75, $p$-value of 0.023). This suggests that AGN-related outflows or emission lines are more morphologically distinct and less aligned with the underlying stellar distribution compared to those in non-AGN galaxies.  \revision{A clear example is 209962 (Figure \ref{fig:example}), where the extended emission forms a bicone-like morphology in the medium-band image.} This could be indicative of the more complex and anisotropic nature of AGN-driven outflows, which can be influenced by the central engine's orientation and the surrounding galactic structure \citep[e.g.,][]{keel_hst_2015}.

The AGN candidates have higher $E(B-V)$ (mean ratio of 1.81 and median ratio of 1.64, $p$-value of 0.002). That is, AGN-hosting galaxies tend to show more dust attenuation, which may obscure the central regions and affect the observed properties of outflows.  \revision{Higher dust attenuation in AGN-hosting galaxies has been widely reported in the literature and is often associated with enhanced star formation and increased gas content \citep[e.g.,][]{netzer_revisiting_2015, ricci_growing_2017, shangguan_gas_2018}. These conditions can contribute to the complexity of the observed emission features and the morphology of outflows.}

The distinct differences in $S_{\rm excess}$, Hausdorff distance, and dust attenuation between AGN and non-AGN candidates highlight the significant role AGN play in driving outflows. AGN-driven outflows are typically more powerful and can impact larger areas compared to those driven by star formation alone. The higher Hausdorff distances for AGN candidates suggest that these outflows are more anisotropic, often presenting as bipolar or complex structures influenced by the AGN's orientation and interaction with the galactic environment. In contrast, outflows in non-AGN galaxies, driven primarily by stellar feedback, are likely more isotropic and aligned with the star-forming regions.

\subsection{Spectroscopic observations}\label{sec:spectra}

\begin{figure*}[!ht]
    \centering
    \includegraphics[width=1\textwidth]{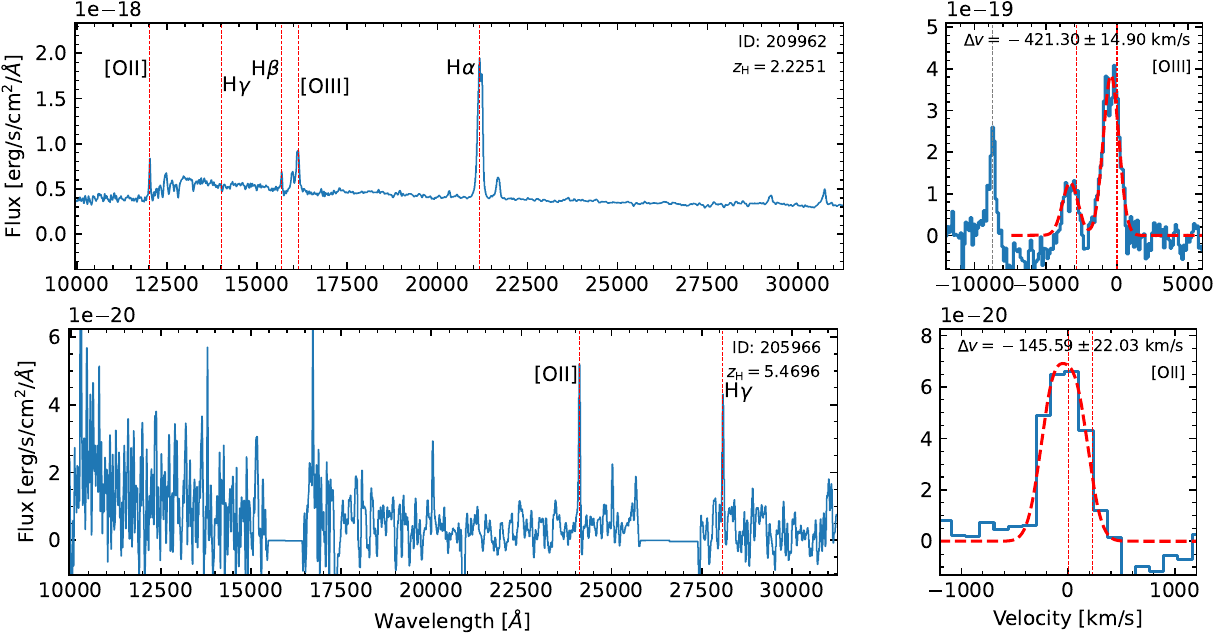}
    \caption{MSA spectra for 2 of our candidates with observations in JADES and SMILES. In each row, the left panel shows the medium resolution spectra, and the fit to the [O III] 4960,5008$\lambda\lambda$ or [O II] 3727,3729$\lambda\lambda$ is shown in the right-hand-side panel. We label the velocity offset of [O III] or [O II] relative to $z_{\rm H}$ for each galaxy. The nominal line centers (with no velocity offset) of the [O III] or [O II] doublets are marked with vertical dashed lines.
    The complete figure set (13 spectra) is presented in the Appendix \ref{app:spectra}.}
    \label{fig:spectra}
\end{figure*}

\begin{figure*}[!ht]
    \centering
    \includegraphics[width=0.7\textwidth]{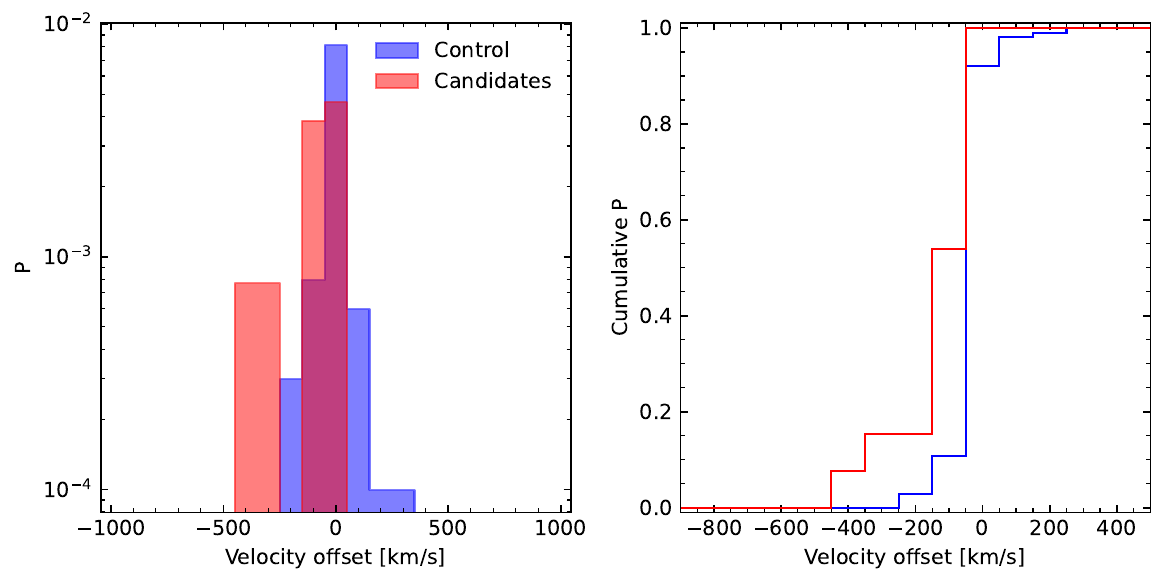}
    \caption{\cut{Probability}  \revision{D}ensity distribution (left) and the corresponding cumulative distribution (right) of velocity offset between [O III] or [O II] emission lines and $z_{\rm H}$. Red and blue colors represent the outflow candidates and control sample, respectively.}
    \label{fig:voff_hist}
\end{figure*}

% In addition to investigating the morphological and galaxy properties from the imaging observations, we find that 13 of our candidates have available medium resolution ($R\sim 1000$) NIRSpec MSA spectra. 
In addition to investigating the morphological and galaxy properties from the imaging observations, 13 of our candidates have available medium-resolution ($R\sim 1000$) NIRSpec MSA spectra.  \revision{Among them, five are classified as AGNs: 197911, 204595, 206907, 209962, and 212228.}
We measure their redshifts, $z_{H}$, based on the S/N-weighted mean of the redshift of H$\alpha$, H$\beta$, and H$\gamma$ lines, if present. Then we refine the redshift determination with visual inspection  \revision{by multiple team members following e.g. \citet{deugenio_jades_2024}. Visual inspection is necessary for almost all candidates due to the presence of outflows, which can shift emission-line centroids or create complex line profiles.} 

Figure \ref{fig:spectra} displays the medium-resolution spectra for these candidates with redshift labeled. The left panels show the overall spectra, highlighting prominent emission lines such as [O III], [O II], and Balmer lines, while the right panels provide detailed fits to these lines.  \revision{Although outflow candidates were selected based on extended H$\alpha$ or [O III] emission in medium-band images, the NIRSpec wavelength coverage limits the available emission lines. As a result, we primarily use ionized oxygen lines ([O III] and [O II]) in spectra to investigate velocity offsets and potential ionized outflow features.}  

The velocity offset ($\Delta v$) of the emission lines relative to $z_{\rm H}$ is also indicated for each galaxy. These velocity offsets can provide a rough indication of the gas kinematics, \cut{helping to differentiate between outflows (blueshift in the emission lines), inflows (redshift in the emission lines), and other dynamic processes.}  \revision{which can be associated with various dynamic processes, including outflows, inflows, and turbulent motions.} An important caveat is that due to the complexity of gas kinematics within a galaxy, $z_{\rm H}$ may not be consistent with the systemic redshift or the emission-line redshift without outflows.

To further analyze the velocity offsets, Figure \ref{fig:voff_hist} shows the \cut{probability} density distribution and the corresponding cumulative distribution of the velocity offsets for the candidates (in red) and the control sample (in blue).  \revision{The control sample consists of galaxies  with JADES NIRSpec observations \citep{deugenio_jades_2024} that do not exhibit significant extended emission-line features, selected to match the redshift within $\Delta z = 0.2$ and the continuum-band flux within 0.3 dex of our candidates.}
The distribution in Figure \ref{fig:voff_hist} indicates that the candidates tend to have greater negative velocity offsets (median $\Delta v=-95\,\rm km\,s^{-1}$) compared to the control sample (median $\Delta v=-5\,\rm km\,s^{-1}$). Such velocity shifts extend beyond the generally symmetric distribution in $\Delta v$ as shown by the control sample. The results potentially \cut{indicate}  \revision{confirm} the presence of galactic-scale outflows. 

We caution that many spectra do not show broad [O III] or [O II] components, which are also important probes of outflows.  \revision{In our sample, the median emission line width is $296 \rm \, km\,s^{-1}$ (FWHM), with only two galaxies (209962, 197911) exhibiting broad components ($> 600 \rm \, km\,s^{-1}$). In both cases, the slits are perpendicular to the extended emission seen in the medium-band images, which may favor the detection of velocity gradients. However, these two galaxies are AGNs, so the broad components may not be solely outflow-driven; alternatively, their outflows might be intrinsically stronger than the non-AGN sample \citep[e.g.,][]{leung_mosdef_2019}.} In other cases, the extended emission is more diffuse with no clear orientation relative to the slit.  \revision{Moreover, as the slits primarily cover the central regions rather than the faint outskirts, this may further limit the detection of broad components.} Therefore, high-resolution integral field spectroscopy is required to confirm the nature of these features.

\section{Summary}\label{sec:summary}

In this work, we conduct the first systematic search to identify and characterize galaxies with potential outflow and/or extended emission line features using medium-band images from the JWST Advanced Deep Extragalactic Survey (JADES) in the GOODS-S field. Our data is featured by deep and high-spatial-resolution NIRCam imaging data and complementary NIRSpec medium resolution spectra for a subset
% include both deep NIRCam imaging data and available NIRSpec medium resolution spectra, 
providing a detailed examination of various galaxy properties. The key results of this paper are summarized below:

\begin{itemize}
    % \item We identified 326 galaxies with potential outflow and/or extended emission line features at \cut{$1 < z < 6$}  \revision{$1.4 < z < 8.4$}. They show significant differences in the medium band morphology compared to the reference bands tracing the stellar continuum.
    \item  \revision{We identified 326 galaxies that show significant differences in the medium band morphology compared to the reference bands tracing the stellar continuum. We attribute these differences to potential outflow and/or extended emission line features at $1.4 < z < 8.4$.} 
    \item Our candidates exhibit significantly higher specific star formation rates, ionization parameters, ionizing photon production efficiencies ($\xi_{\rm{ion}}$), and ionizing photon production rates ($\dot{n}_{\rm{ion}}$) compared to a control sample, suggesting more active and dynamic star-forming environments.
    \item The extent of potential outflows or extended emission lines shows strong correlations with SFR and $\dot{n}_{\rm{ion}}$, highlighting the critical role of star formation and ionizing photon production in driving these features.
    \item Galaxies hosting AGN tend to have larger $S_{\rm excess}$, higher Hausdorff distances, and greater dust attenuation compared to non-AGN galaxies, indicating that AGN-driven outflows are more extensive, non-isometric, and associated with more dust.
    \item Spectroscopic observations of 13 candidates with medium resolution NIRSpec MSA spectra suggest significant velocity offsets in emission lines such as [O III] and [O II], providing further evidence of potential dynamic outflow activity.
\end{itemize}

Our work highlights the robustness of deep medium-band imaging observations in studying galaxy evolution. The candidate galaxy sample will provide important legacy value in identifying and characterizing outflow candidates across cosmic time. Future spectroscopic follow-up on our sample will enhance our understanding of the origins and characteristics of galactic outflows and provide valuable insights into the complex interactions between supermassive black holes and their host galaxies.

\section*{Acknowledgments}
% \begin{acknowledgments}
 \revision{We would like to thank the anonymous reviewer for their constructive feedback.}
YZ, MJR, ZJ, PAC, BR, and CNAW are supported by JWST/NIRCam contract to the University of Arizona NAS5-02015. BR also acknowledges support from JWST program 3215.
CS acknowledges support from the Science and Technology Facilities Council (STFC), by the ERC through Advanced Grant 695671 ``QUENCH'', by the UKRI Frontier Research grant RISEandFALL. 
SA and GHR acknowledge support from the JWST Mid-Infrared Instrument (MIRI) Science Team Lead, grant 80NSSC18K0555, from NASA Goddard Space Flight Center to the University of Arizona.
AJB and GCJ acknowledge funding from the ``FirstGalaxies'' Advanced Grant from the European Research Council (ERC) under the European Union’s Horizon 2020 research and innovation programme (Grant agreement No. 789056).
SC acknowledges support by European Union’s HE ERC Starting Grant No. 101040227 - WINGS.
JS acknowledges support by the Science and Technology Facilities Council (STFC), ERC Advanced Grant 695671 ``QUENCH''.
H{\"U} gratefully acknowledges support by the Isaac Newton Trust and by the Kavli Foundation through a Newton-Kavli Junior Fellowship.
The research of CCW is supported by NOIRLab, which is managed by the Association of Universities for Research in Astronomy (AURA) under a cooperative agreement with the National Science Foundation.

This work is based on observations made with the NASA/ESA/CSA James Webb Space Telescope. The data were obtained from the Mikulski Archive for Space Telescopes at the Space Telescope Science Institute, which is operated by the Association of Universities for Research in Astronomy, Inc., under NASA contract NAS 5-03127 for JWST. These observations are associated with PID 1180, 1207, 1210, 1286, and 1963. The specific observations analyzed can be accessed via \dataset[https://doi.org/10.17909/8tdj-8n28]{https://doi.org/10.17909/8tdj-8n28},  \dataset[https://dx.doi.org/10.17909/fsc4-dt61]{https://dx.doi.org/10.17909/fsc4-dt61}, and \dataset[https://doi.org/10.17909/et3f-zd57]{https://doi.org/10.17909/et3f-zd57}.

The authors acknowledge use of the lux supercomputer at UC Santa Cruz, funded by NSF MRI grant AST 1828315.

We respectfully acknowledge the University of Arizona is on the land and territories of Indigenous peoples. Today, Arizona is home to 22 federally recognized tribes, with Tucson being home to the O'odham and the Yaqui. Committed to diversity and inclusion, the University strives to build sustainable relationships with sovereign Native Nations and Indigenous communities through education offerings, partnerships, and community service.
% \end{acknowledgments}

\vspace{5mm}
\facilities{JWST, MAST}

\software{
{\tt Astropy} \citep{astropy_collaboration_astropy_2013,astropy_collaboration_astropy_2018,astropy_collaboration_astropy_2022},
{\tt FitsMap} \citep{hausen_fitsmap_2022},
{\tt JWST Calibration Pipeline} \citep{bushouse_jwst_2022},
{\tt scikit-learn} \citep{pedregosa_scikit-learn_2011}
}

\appendix

\section{Galaxy Properties Inferred with Different SED Parameters}\label{app:simmonds}
Here, we compare our SED fitting results to those in \citet{simmonds_ionising_2024}, which use different priors and assumptions. \citet{simmonds_ionising_2024} fit the SED using {\tt Prospector} for $z>3$ galaxies in JADES GOODS-S, following a similar procedure as described in Section \ref{sec:data}. They also use a non-parametric star formation history (SFH) described by \citet{leja_how_2019}. However, their SFH is modeled as eight SFR bins controlled by the bursty-continuity prior \citep{tacchella_stellar_2022}. In addition, they adopt a Chabrier \citep{chabrier_galactic_2003} IMF instead of the Kroupa \citep{kroupa_variation_2001} IMF.

\begin{figure*}[!ht]
    \centering
    \includegraphics[width=1\textwidth]{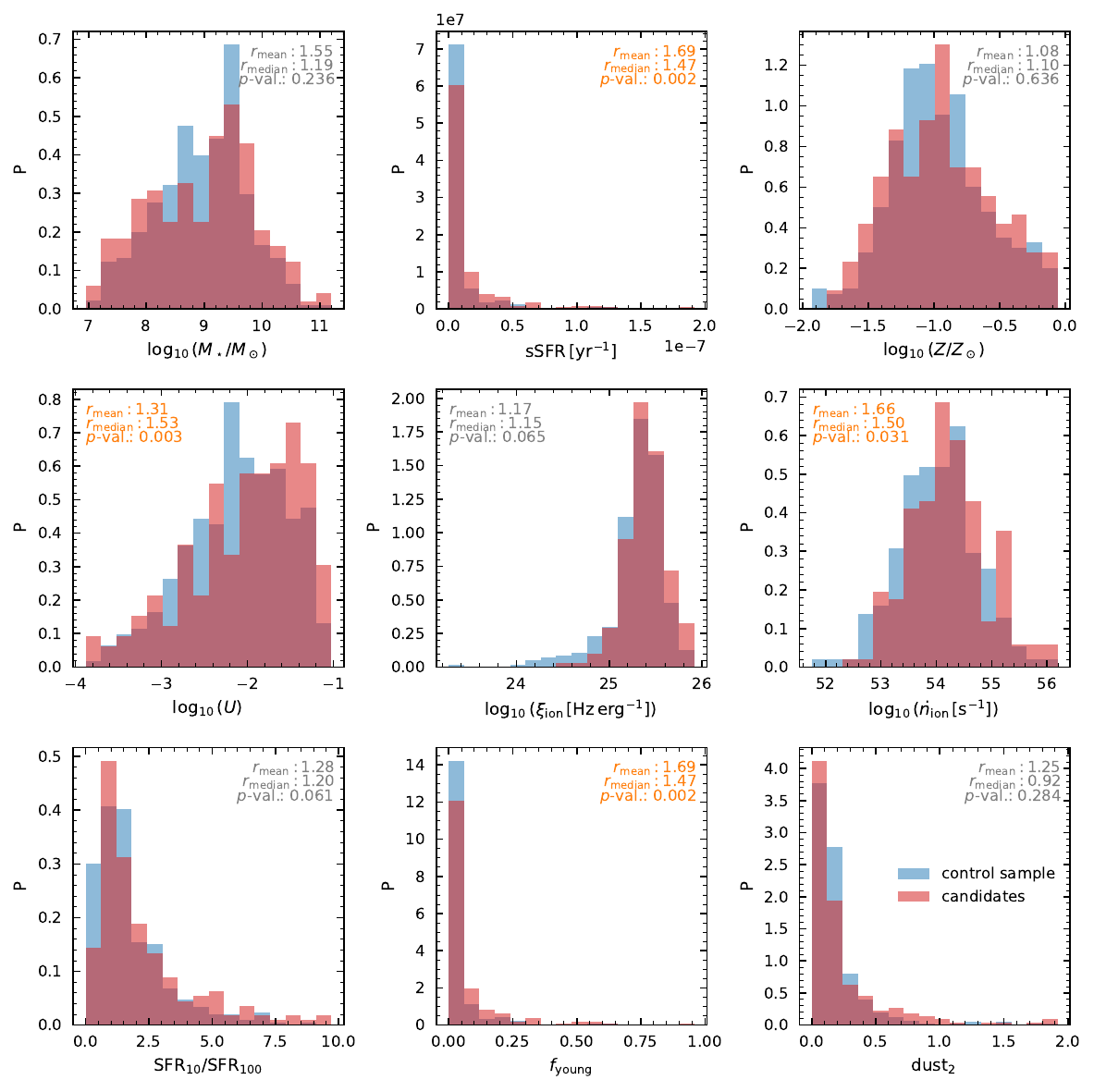}
    \caption{Similar to Figure \ref{fig:candidates_vs_control}, but showing results based on galaxy properties in \citet{simmonds_ionising_2024} for our $z>3$ galaxies. As in Figure \ref{fig:candidates_vs_control}, we \cut{only label the plot}  \revision{highlight the labels} when there is a significant difference between the two distributions.}
    \label{fig:candidates_vs_controlc}
\end{figure*}

\begin{figure*}[!ht]
    \centering
    \includegraphics[width=1\textwidth]{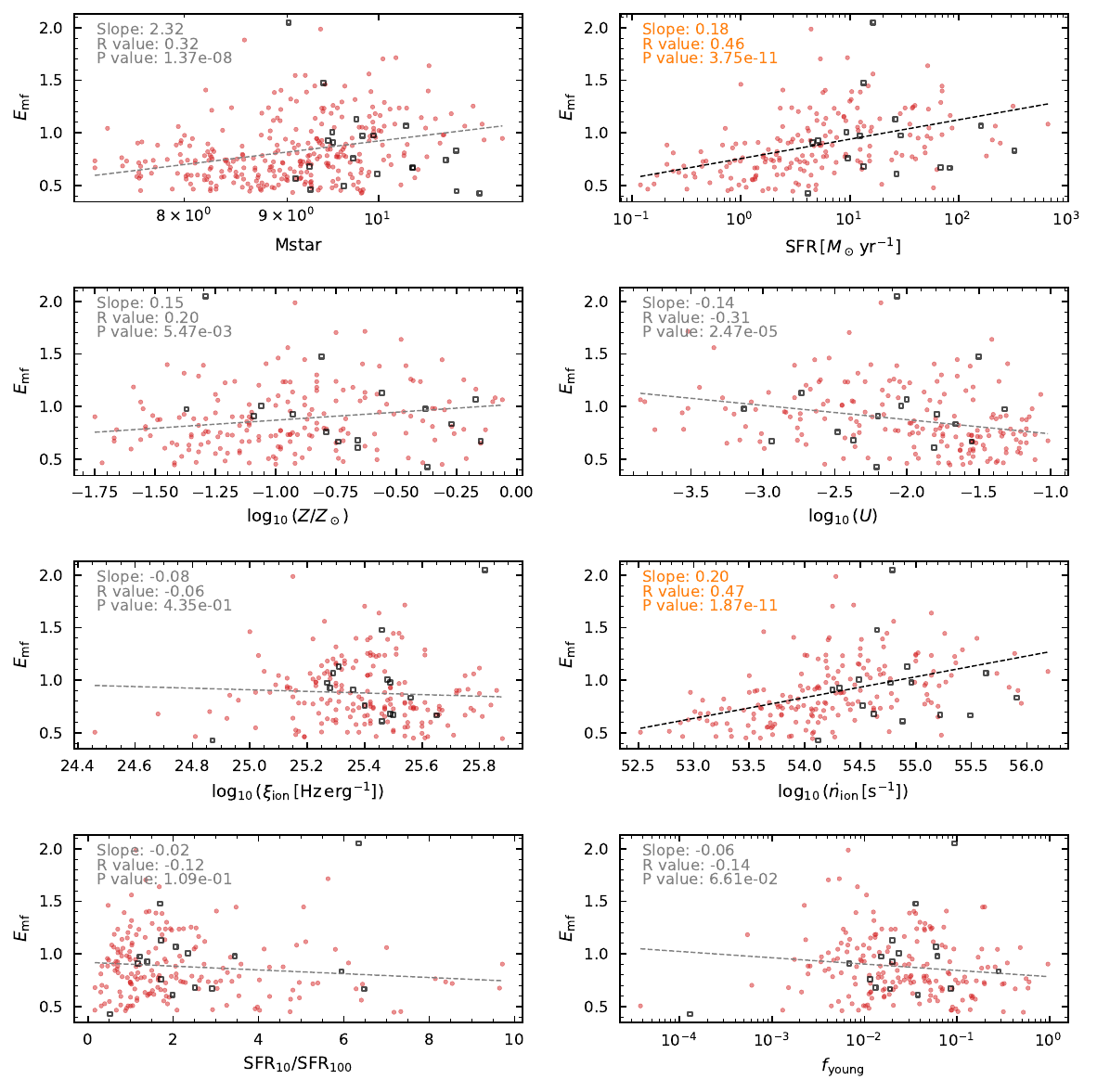}
    \caption{Similar to Figure \ref{fig:ext_vs_parameters}, but showing results based on galaxy properties in \citet{simmonds_ionising_2024} for our $z>3$ galaxies.}
    \label{fig:ext_vs_parametersc}
\end{figure*}

Figures \ref{fig:candidates_vs_controlc} \& \ref{fig:ext_vs_parametersc} show that our results do not change significantly if we use the galaxy properties in \citet{simmonds_ionising_2024}. We highlight that \citet{simmonds_ionising_2024} also measure the ionizing photon production efficiency ($\xi_{\rm ion}$) and the ionizing photon production rate ($\dot{n}_{\rm ion}$). The total ionizing photon budget is consistent with the recent cosmic reionization history measured based on quasar absorption lines \citep[e.g.,][]{becker_new_2013, gaikwad_measuring_2023,jin_nearly_2023,zhu_damping_2024} and potentially relieves the tension indicated for the total ionizing photon budget during reionization \citep[e.g.,][]{munoz_reionization_2024}.

\section{MSA Spectra for Candidates with Potential Outflows}
\label{app:spectra}

In this appendix, we present the full set of 13 MSA spectra in Figure \ref{fig:spectra0}. Spectra with IDs 209962, 205966, 204595, 212228, and 219050 are from the SMILES program \citep[][and Y.~Zhu et al.~in prep.]{alberts_smiles_2024,rieke_smiles_2024}, and the rest of the spectra are from the JADES NIRSpec data release \citep{deugenio_jades_2024}.

\begin{figure*}
    \centering
    \includegraphics[width=0.7\textwidth]{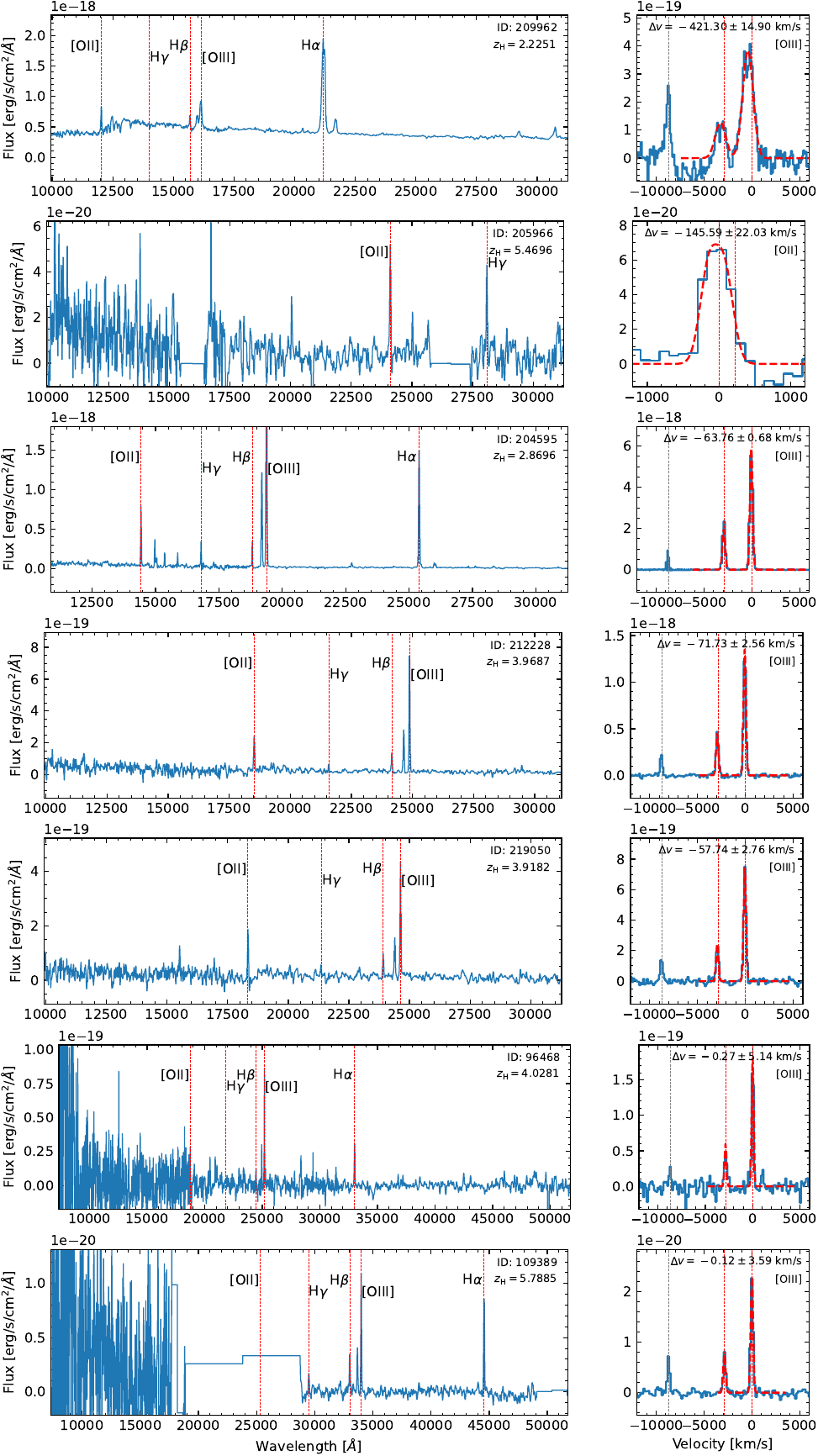}
    \caption{MSA spectra for 13 of our candidates with observations in JADES and SMILES. In each row, the left panel shows the medium resolution spectra, and the fit to the [O III] 4960,5008$\lambda\lambda$ or [O II] 3727,3729$\lambda\lambda$ is shown in the right-hand-side panel. We label the velocity offset of [O III] or [O II] relative to $z_{\rm H}$ for each galaxy. The nominal line centers (with no velocity offset) of the [O III] or [O II] doublets are marked with vertical dashed lines. (to be continued).}
    \label{fig:spectra0}
\end{figure*}

\begin{figure*}
    \centering
    \includegraphics[width=0.7\textwidth]{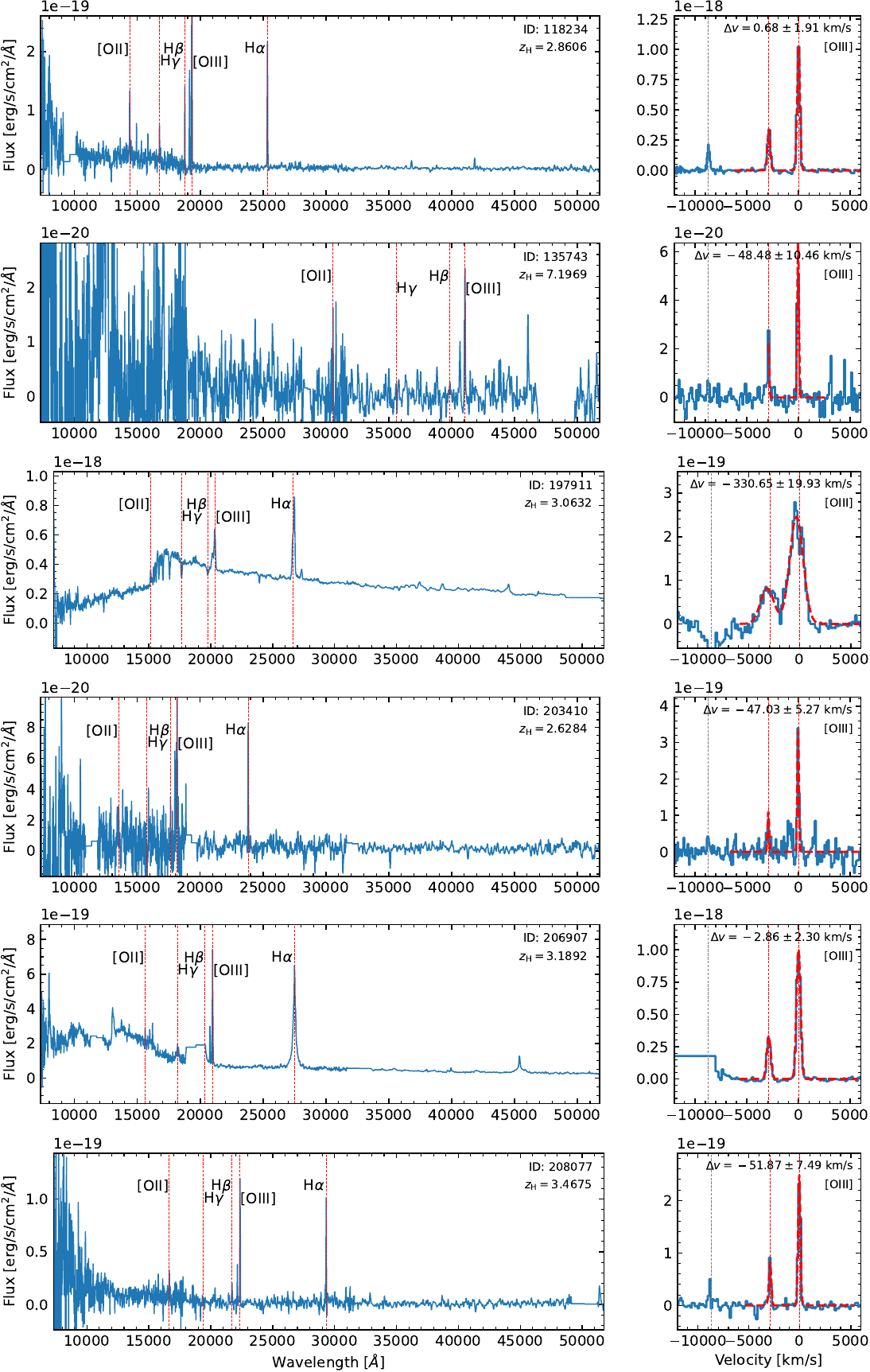}
    \addtocounter{figure}{-1}
    \caption{Continued.}
    \label{fig:spectra1}
\end{figure*}

\clearpage

%\bibliography{outflows}{}
%\bibliographystyle{aasjournal}

%\allauthors
%\listofchanges

\end{document}